\documentclass[11pt,a4,twoside]{article}
\usepackage[dvips,a4paper,left=2.60cm,right=2.60cm,top=2.98cm,bottom=2.98cm,headsep=1em]{geometry}
\usepackage{fancyhdr}
\usepackage{titlesec}
\usepackage[latin1]{inputenc}
\usepackage{amsmath,amsfonts,amssymb,amstext}
\usepackage{graphicx}
\usepackage{subcaption}
\usepackage{rotating}
\usepackage[font={small}]{caption}
\usepackage{natbib}
\usepackage{lmodern,slantsc}
\usepackage{relsize}
\usepackage[breaklinks, colorlinks=true, linkcolor=blue,
citecolor=black, urlcolor=blue, pdfborder={0 0 0},
pdfpagelabels]{hyperref}
\usepackage{tikz}
\DeclareMathAlphabet{\mathsfbf}{OT1}{cmss}{bx}{n}
\DeclareMathAlphabet{\mathmibf}{OT1}{cmr}{bx}{it}
\DeclareMathAlphabet\mathbfcal{OMS}{cmsy}{b}{n}
\def\vx{\mathbf x}
\def\vy{\mathbf y}
\def\vz{\mathbf z}

\def\vu{\mathbf u}

\def\vz{\mathbf z}

\def\vA{\mathbf A}

\def\vH{\mathbf H}

\def\vK{\mathbf K}

\def\v0{\boldsymbol{0}}

\def\vrho{\boldsymbol{\rho}}

\def\L{\langle}
\def\R{\rangle}
\newlength{\FigureHeight}
\newlength{\FigureHeightHalf}

\numberwithin{equation}{section}
\titleformat{\section}
{\large\bfseries}{\thetitle.}{0.5em}{}
\titleformat{\subsection}
{\normalfont\itshape\filcenter}{\normalfont\thetitle.}{0.5em}{}
\begin{document}

\title{\vspace{-1em} An example elucidating the mathematical situation in the statistical
non-uniqueness problem of turbulence}
\author{Michael Frewer\thanks{Email address for correspondence:
frewer.science@gmail.com}\\
\small Tr\"ubnerstr. 42, 69121 Heidelberg, Germany}
\date{{\small\today}}
\clearpage \maketitle \thispagestyle{empty}

\vspace{-2em}
\begin{abstract}
\noindent An instructive example is presented to elucidate the
mathematical situation in the non-uniqueness problem of the
infinite Friedmann-Keller hierarchy of equations for all
multi-point moments within the theory of spatially unbounded
Navier-Stokes turbulence. It is shown that the non-uniqueness
problem of the Friedmann-Keller hierarchy emerges from the
property that the system of equations is defined forward
recursively. As a result, this system does not possess a unique
general solution, even when the complete infinite system is
formally considered. That is, even when imposing a sufficient
number of initial conditions to this infinite system, it still
does not provide a unique solution. This finding is supported by a
Lie-group invariance analysis, in that the imposed example
analogous to the Friedmann-Keller hierarchy admits an unclosed Lie
algebra which allows for infinitely many functionally different
equivalence transformations which all can be made compatible with
any specifically chosen initial condition. Hence, if no prior
modelling assumptions are made to close the Friedmann-Keller
system of equations, the existence of an invariant solution within
such a forward recursively defined system is then without value,
since it just represents an arbitrary solution among infinitely
many other, equally privileged invariant~solutions.

\vspace{0.5em}\noindent{\footnotesize{\bf Keywords:} {\it
Turbulence, Hopf Equation, Friedmann-Keller Hierarchy, Multi-point
Correlations, Functional Equations, Ordinary and Partial
Differential Equations, Infinite Systems, Lie Groups, Lie Algebra,
Symmetries and Equivalences, Unclosed Systems, General Solutions,
Cauchy~Problems}}
\end{abstract}

\section{Introduction and motivation}

Due to the non-linear evolution of the Navier-Stokes equations,
the statistical description of a turbulent flow on the level of
the velocity moments is inherently unclosed. Presumably the first
attempt to systematically formulate the equations of moments for
turbulence was given by the proposed method of
\cite{Keller24}\footnote[2]{For a more concise derivation of the
Friedmann-Keller equations, see e.g. \cite{Monin71}.}. The result
is an infinite hierarchy of coupled multi-point correlation (MPC)
equations. Theoretical considerations show that on the level of
the moments all (infinite) correlation orders have to be taken
into account in order to allow for a consistent statistical
description of turbulence, which, since then, became known as the
still prevailing closure problem of turbulence. The motivation of
\cite{Hopf52} was to formally bypass this problem in formulating a
statistical description which operates on a single closed
equation: Known as the functional Hopf equation\footnote[3]{Note
that also the discrete version of the Hopf equation, the
Lundgren-Monin-Novikov hierarchy, which constitutes an infinite
chain of coupled integro-differential equations, remains to be a
closed system. The reason is that next to any possible initial
conditions sufficient internal integral conditions are defined
such that its general solution manifold is unique. For more
details, see e.g. \cite{Hosokawa06} along with Appendix C in
\cite{Frewer14.1} and the references therein.}, it then induces
the unclosed infinite hierarchy of MPC equations through a single
moment generating functional~$\Phi$.

Although one gained through this extension the theoretical
advantage of dealing with a formally closed statistical
description, the practical situation of constructing specific
solutions of the problem, however, did not improve substantially
when compared to the unclosed description of the moments. The fact
is that on both levels of statistical description we face serious
drawbacks: On the higher, more abstract level of the Hopf equation
we face the problem of dealing with a functional calculus which
from the outset is difficult to access numerically as well as
analytically in a practical and constructive sense, while on the
lower level of the Friedmann-Keller equations we face the problem
of dealing with a hierarchy of coupled partial differential
equations (PDEs) which is infinite. As I will show in this study
by analyzing a lower-dimensional analogue, it becomes apparent
that the latter problem faces a yet more fundamental problem: Even
when formally considering all (infinite) equations in the
statistical hierarchy, i.e. even if the Friedmann-Keller system is
{\it not} truncated, the full system itself is still unclosed
since no unique {\it general} solution can be constructed. In
other words, when considering the case of a spatially unbounded
flow e.g. in stating it mathematically as a Cauchy problem (as
will be done throughout this study), the infinite system still is
underdetermined and thus unclosed in that it does not offer any
unique solutions although specific initial conditions have been
posed. The reason for this defect is discussed in this study at
the example of a lower-dimensional analogue, which shows all
primary features of the Hopf-induced Friedmann-Keller hierarchy of
MPC equations when stated as a functional Cauchy problem. These
features are:
\begin{itemize}
\item[i)] the linearity of the higher level equation (the Hopf
equation is a linear equation),
\item[ii)] the linearity of the induced lower level moment
equations (the infinite Friedmann-Keller hierarchy is a linear
system),
\item[iii)] the infinite hierarchy of moments is ordered forward
recursively (each equation in the Friedmann-Keller hierarchy
contains a higher order correlation function which only enters the
next higher order equation in this chain).
\end{itemize}
The non-uniqueness problem just mentioned above is rooted solely
in feature iii), in that a certain hierarchy of differential
equations is defined {\it forward} recursively. Because, as I will
demonstrate, if the infinite chain of equations would be ordered
oppositely, namely {\it backward} recursively (where each equation
contains a lower order correlation function which then only enters
the next lower order equation in this chain), then the
non-uniqueness problem does not exist. But since the infinite
Friedmann-Keller chain of moments is defined as a forward
recursive system, it inherently faces the problem of
non-uniqueness in its {\it general} solution manifold even if all
infinite equations of the system are formally taken along.
Consequently, when performing an invariance analysis on such a
system, one can generate infinitely many and functionally diverse
Lie-point symmetries\footnote[2]{In order to warrant consistency
within an invariance analysis as for the Friedmann-Keller system,
the actual symmetries must be identified as weaker equivalences.
The reason is that since the considered
system~of~equations\linebreak is unclosed, all admitted
invariances only act as equivalence transformations and not as
symmetry trans\-formations. For more details, see e.g. Section 2
in \cite{Frewer14.1}, Sections 3-5 in~\cite{Frewer14.3}, and
the~references therein.}, and hence infinitely many different and
independent invariant solutions which all can be made compatible
to any posed initial condition.

My motivation for this study was to elucidate this non-uniqueness
problem for the MPC equations in all its facets, because it seems
that in the relevant literature on turbulence there still exists a
misconception on this issue, in particular in the studies of
Oberlack~et~al., e.g. in \cite{Oberlack10}, \cite{Oberlack13.1},
\cite{Oberlack14}, \cite{Oberlack14.1}, and \cite{Oberlack14.2}.
In all these studies the reduced (lower level) Friedmann-Keller
hierarchy of multi-point moments is incorrectly treated as a {\it
closed} system, with the misleading argument that it's due to that
the system is infinite dimensional, because "...~if the infinite
set of correlation equations is considered the closure problem is
somewhat bypassed" \cite[p.~454]{Oberlack10}. Also the performed
invariance analysis in each of these studies is misleading. For
example, to obtain the result that "...~the latter set of
symmetries is considerably enlarged for the infinite set of MPC
equations" \cite[p.~12]{Oberlack13.1}\footnote[2]{In
\cite{Frewer14.1,Frewer14.3} it has been shown that this newly
enlarged set of symmetries is even unphysical in that it leads to
various inconsistencies with the underlying deterministic
Navier-Stokes theory. Hence, besides the well-known classical
symmetries of the Euler and Navier-Stokes equations
\citep{Fushchich93,Olver93,Frisch95,Andreev98}, no new enlarged
physical set of statistical symmetries exist.} is not a surprising
result, because the considered infinite Friedmann-Keller system of
moments, as we will assert herein due to its forward recursively
organized hierarchy, is simply an unclosed and thus an
underdetermined system of equations which intrinsically allows for
arbitrary invariances. Furthermore, to address the open question
that "...~so far completeness of all admitted symmetries of the
MPC equation has not been shown... [which] appears to be necessary
not only from a theoretical point of view but rather essential to
generate scaling for all higher moments"
\cite[p.~1702]{Oberlack14.2}, and that "...~hence, finding new
statistical symmetries and/or proving the completeness of the set
of symmetries is a next task for further study"
\cite[p.~10]{Oberlack14.1}, would collectively only lead to a
superfluous task, because, as already said, arbitrary invariances
can be constructed when considering an unclosed infinite hierarchy
of equations as the MPC equations considered by Oberlack~et~al.;
therefore completeness in this very sense can never be obtained in
such infinite~systems.\linebreak Finally, when trying to correctly
interpret any results arising from such an invariance analysis, in
particular when trying to construct invariant solutions for such
unclosed systems, it is necessary to recognize the subtle but
important difference between a symmetry transformation and an
equivalence transformation. While a symmetry transformation always
maps a solution to another solution of the same (closed) equation,
an equivalence transformation, in contrast, generally only maps a
possible solution of one underdetermined (unclosed) equation to a
possible solution of another underdetermined (unclosed) equation
(see e.g. \cite{Ovsiannikov82}, \cite{Meleshko96},
\cite{Ibragimov04}, \cite{Bila11}). Hence, since any Lie-group
invariance analysis on the infinite Friedmann-Keller hierarchy
will be based on equivalence and not on symmetry groups, it is
misleading and even ill-defined to construct invariant solutions
of the\linebreak unclosed Friedmann-Keller equations if no
modelling assumptions on these set of equations are being priorly
invoked~\citep{Frewer14.1,Frewer14.3}. Therefore, to state that
"...~a variety of classical\linebreak and new scaling laws" were
derived for which "it was shown that they are exact solutions of
symmetry invariant type of the infinite dimensional series of MPC
equations" \cite[p.~102]{Oberlack14} is more than misleading,
because what to understand under "exact solution"\linebreak when
the construction process for generating solutions itself is
arbitrary? Independent of whether a Lie group invariance analysis
is employed or not, these "exact solutions" are nothing else but
arbitrary and thus non-privileged solutions. They just arise from
an {\it unclosed} infinite dimensional system which does not give
any indication of whether these "exact results" are relevant
solutions or not. For that, additional, {\it external} information
is needed, e.g. as posing certain modelling assumptions to close
the infinite system of equations.

The following study is organized as follows: Section 2 introduces
the functional Hopf equation in physical space, from which the
infinite Friedmann-Keller system of multi-point moments for
spatially unconfined turbulence is then reduced. Section 3
investigates a lower-dimensional analogue to the system
established in Section 2, but, which first will only feature the
properties i) and ii) listed above. Instead of property iii), a
backward recursively infinite system is first considered, in order
to demonstrate that for such a case the non-uniqueness problem as
discussed above cannot arise. Section 4 forms the key part of this
study, in that an extended lower-dimensional analogue is
considered which now will feature {\it all} properties i)-iii) of
the system established in Section~2. This analogue will allow us
to elucidate step by step all facets of the non-uniqueness problem
which inherently characterizes the infinite Friedmann-Keller
system of multi-point moments. Section 5 concludes and completes
the investigation.

\section{The Hopf equation and its induced MPC equations}

The functional Hopf equation \citep{Hopf52,McComb90} for the
characteristic or moment generating functional
$\Phi=\Phi[\vy(\vx),t]$ in physical space is given by the {\it
linear} differential equation
\begin{equation}
\frac{\partial \Phi}{\partial t} =\int y^{\perp}_\alpha(\vx)
\left(i\frac{\partial}{\partial x_\beta}\frac{\delta^2
\Phi}{\delta y_\alpha(\vx)\delta
y_\beta(\vx)}+\nu\Delta\frac{\delta\Phi}{\delta
y_\alpha(\vx)}\right)d^3\vx, \label{150809:1945}
\end{equation}
where $\vy(\vx)$ is a real, integrable and time-independent
auxiliary vector field being complementary to the ensemble of all
admissible incompressible velocity fields $\vu(\vx)$ governed by
the Navier-Stokes equations sampled at each time step $t\geq 0$.
Since we consider an unbounded flow, we assume that the ensemble
of fields $\vy(\vx)$ is vanishing sufficiently fast at infinity.
The fields~$\vy^{\perp}$ in equation \eqref{150809:1945} represent
the transverse (solenoidal) part of $\vy(\vx)$ in order to
eliminate the pressure terms which otherwise would arise in
\eqref{150809:1945}. We will use the transverse form as used in
\cite{Hopf52} (see also Appendix A)
\begin{equation}
y^{\perp}_\alpha(\vx)=y_\alpha(\vx)-\partial_\alpha\varphi(\vx),
\label{150809:2110}
\end{equation}
with the scalar field
\begin{equation}
\varphi(\vx)=-\int\frac{\nabla^\prime\cdot
\vy(\vx^\prime)}{4\pi|\vx-\vx^\prime|}\, d^3\vx^\prime.
\end{equation}
Note that in order to warrant physical consistency, a mathematical
solution of the Hopf equation \eqref{150809:1945} is only admitted
if for all times the following conditions for the characteristic
functional are fulfilled
\begin{equation}
\Phi^*[\vy(\vx),t]=\Phi[-\vy(\vx),t],\quad\;\;\;
\Phi[0,t]=1,\quad\;\;\;\big\vert\Phi[\vy(\vx),t]\big\vert\leq 1.
\label{140124:2059}
\end{equation}
Now, when reducing the MPC equations from \eqref{150809:1945}, one
can either work in the re\-pre\-sen\-tation of the full vector
field $\vy$, or in the decomposed representation of its transverse
part $\vy^\perp$. The former representation of the Hopf equation
\eqref{150809:1945} takes the form (see Appendix B.1)
\begin{align}
\frac{\partial \Phi}{\partial t} & = \int y_\alpha(\vx)
\left(i\frac{\partial}{\partial x_\beta}\frac{\delta^2
\Phi}{\delta y_\alpha(\vx)\delta
y_\beta(\vx)}+\nu\Delta\frac{\delta\Phi}{\delta
y_\alpha(\vx)}\right)d^3\vx\nonumber\\
&\qquad\qquad\quad +  i\!\int
y_\alpha(\vx)\left(\frac{\partial}{\partial
x_\alpha}\frac{1}{4\pi|\vx-\vx^\prime|}\right)\frac{\partial^2}{\partial
x^\prime_\beta\partial x^\prime_\gamma}\frac{\delta^2 \Phi}{\delta
y_\beta(\vx^\prime)\delta y_\gamma(\vx^\prime)}d^3\vx^\prime\,
d^3\vx,
\label{150811:1222}
\end{align}
while the latter one takes the more concise form (see Appendix
B.2)
\begin{align}
\frac{\partial \Phi}{\partial t} & = \int y^{\perp}_\alpha(\vx)
\left(i\frac{\partial}{\partial x_\beta}\frac{\delta^2
\Phi}{\delta y^\perp_\alpha(\vx)\delta
y^\perp_\beta(\vx)}+\nu\Delta\frac{\delta\Phi}{\delta
y^\perp_\alpha(\vx)}\right)d^3\vx.
\label{150811:1223}
\end{align}
Note that both equations, \eqref{150811:1222} and
\eqref{150811:1223}, refer to the same characteristic functional
$\Phi$, since under the defining relation \eqref{150809:2110} it
transforms invariantly $\Phi[\vy(\vx),t]=\Phi[\vy^\perp(\vx),t]$
\cite[p.~93]{Hopf52}. Keep in mind that equation
\eqref{150811:1223} is a projection of equation
\eqref{150811:1222} onto solenoidal vector fields, where if
$\nabla\cdot \vy\neq 0$ then $\nabla\cdot \vy^\perp = 0$. In other
words, if the ensemble of all fields $\vy$ would be
divergence-free, i.e. $\nabla\cdot \vy= 0$ (which in general is
{\it not} assumed), then \eqref{150811:1222} is identical to
\eqref{150811:1223}, as can be readily observed in
\eqref{150811:1222} after executing a partial integration with
respect to~$\vx$.

Further note that both representations \eqref{150811:1222} and
\eqref{150811:1223} of the Hopf equation \eqref{150809:1945} lead
to the same set of MPC equations. It's obvious that the latter
representation \eqref{150811:1223} is suited best for any formal
investigations as we are interested herein. It induces
considerably simplified formal expressions for the MPC equations
since no explicit pressure terms would surface as in
representation \eqref{150811:1222}. The corresponding
pressure-dependent MPC equations, when needed, can then either be
derived by using the defining relation \eqref{150809:2110} in the
obtained results from the transverse-projected equation
\eqref{150811:1223}, or by directly deriving these from the
full-field equation~\eqref{150811:1222}.

According to \cite[p.~100-101]{Hopf52} the infinite
Friedmann-Keller hierarchy of MPC equations can be generated from
\eqref{150811:1223} by performing a Taylor expansion of
$\Phi[\vy^\perp,t]$ around the functional
point~$\vy^\perp=\boldsymbol{0}$:
\begin{multline}
\Phi[\vy^\perp,t]=1+\int d^3\vx_1\,
y^\perp_{\alpha_1}(\vx_1)\frac{\delta\Phi[\vy^\perp,t]}{\delta
y^\perp_{\alpha_1}(\vx_1)}\bigg\vert_{\vy^\perp=\boldsymbol{0}}\\
+ \frac{1}{2!}\int d^3\vx_1\, d^3\vx_2\,
y^\perp_{\alpha_1}(\vx_1)\,y^\perp_{\alpha_2}(\vx_2)
\frac{\delta^2\Phi[\vy^\perp,t]}{\delta
y^\perp_{\alpha_1}(\vx_1)\delta
y^\perp_{\alpha_2}(\vx_2)}\bigg\vert_{\vy^\perp=\boldsymbol{0}}+\;\;\;\cdots
,\quad \label{150826:1146}
\end{multline}
which formally can be written as
\begin{equation}
\Phi[\vy^\perp,t]=1+\Phi^1[\vy^\perp,t]+\Phi^2[\vy^\perp,t]
+\cdots , \label{150811:2131}
\end{equation}
where $\Phi^n[\vy^\perp,t]$ is a homogeneous polynomial functional
of degree $n$ in $\vy^\perp=\vy^\perp(\vx)$,
\begin{equation}
\Phi^n[\vy^\perp,t]=\int d^3\vx_1\cdots d^3\vx_n\,
y^\perp_{\alpha_1}(\vx_1)\cdots y^\perp_{\alpha_n}(\vx_n)\,
K^\perp_{\alpha_1\dotsc \alpha_n}(\vx_1,\dotsc ,\vx_n ,t),
\label{150812:0812}
\end{equation}
and where the components of the kernel function
$\vK^\perp_n=(K^\perp_{\alpha_1\dotsc \alpha_n})$ are given by
\begin{align}
K^\perp_{\alpha_1\dotsc \alpha_n}(\vx_1,\dotsc ,\vx_n ,t)
&\,\,=\,\, \frac{1}{n!}\, \frac{\delta^n\Phi[\vy^\perp,t]}{\delta
y^\perp_{\alpha_1}(\vx_1)\cdots\delta
y^\perp_{\alpha_n}(\vx_n)}\bigg\vert_{\vy^\perp=\boldsymbol{0}}
\nonumber\\[0.75em]
&\,\,=\,\,\frac{1}{n!}\, \frac{\delta^n\Phi[\vy,t]}{\delta
y_{\alpha_1}(\vx_1)\cdots\delta
y_{\alpha_n}(\vx_n)}\bigg\vert_{\vy=\boldsymbol{0}}\;\: =\,\,
K_{\alpha_1\dotsc \alpha_n}(\vx_1,\dotsc ,\vx_n ,t).
\label{150811:2104}
\end{align}

\noindent The last equality stems from the fact the functional
derivatives stay invariant under the transformation
\eqref{150809:2110} (see Appendix B.2); and since
$\vy^\perp=\boldsymbol{0}\,\Leftrightarrow\, \vy=\boldsymbol{0}$,
the kernel function constitutes itself as an invariant function,
i.e. $\vK^\perp_n=\vK_n$. In \cite{Hopf52} this function is
determined as the equal-time MPC function of the incompressible
Navier-Stokes velocity field~$\vu(\vx,t)$
\begin{equation}
K_{\alpha_1\dotsc \alpha_n}(\vx_1,\dotsc ,\vx_n
,t)=\frac{i^n}{n!}\big\L u_{\alpha_1}(\vx_1,t)\cdots
u_{\alpha_n}(\vx_n,t)\big\R, \label{150811:2353}
\end{equation}
where $\L\cdot\R$ denotes the statistical ensemble average of the
instantaneous and thus fluctuating velocity field evaluated at $n$
different points. Upon inserting the power series of $\Phi$
\eqref{150811:2131} into equation \eqref{150811:1223} and upon
equating terms of equal degree on both sides, one finally obtains
the infinite sequence of differential equations for the MPC
functions
\begin{equation}
\frac{\partial \Phi^n}{\partial t} = \int y^{\perp}_\alpha(\vx)
\left(i\frac{\partial}{\partial x_\beta}\frac{\delta^2
\Phi^{n+1}}{\delta y^\perp_\alpha(\vx)\delta
y^\perp_\beta(\vx)}+\nu\Delta\frac{\delta\Phi^n}{\delta
y^\perp_\alpha(\vx)}\right)d^3\vx, \quad \forall n\geq 1.
\label{150812:0807}
\end{equation}

\noindent By taking the notation from \cite{Oberlack10}, the above
system can be evaluated~to (see Appendix C.1)
\begin{equation}
\frac{\partial \vH^\perp_n}{\partial t} + \sum_{i=1}^n
\nabla_{\vx_i}\cdot
\widehat{\vH}^\perp_{i,n+1}-\nu\sum_{i=1}^n\Delta_{\vx_i}\vH_n^\perp
=\boldsymbol{0}, \quad \forall n\geq 1, \label{150817:1647}
\end{equation}
where
\begin{equation}
\vH^\perp_n(\vx_1,\dotsc,\vx_n,t)=\vH_n(\vx_1,\dotsc,\vx_n,t)=
\big\L \vu(\vx_1,t)\otimes\dots\otimes\vu(\vx,t)\big\R,
\end{equation}
is the instantaneous (equal-time) multi-point velocity correlation
function of $n$-th order, and where
\begin{equation}
\!\widehat{\vH}^\perp_{i,n+1}(\vx_1,\dotsc,\vx_i,\dotsc\vx_n,\vx_i,t)
=\left(\lim_{\vx_{n+1}\to\vx_i}
=\vH_{n+1}(\vx_1,\dotsc,\vx_n,\vx_{n+1},t) \right)^\perp \!
,\;\:\,\, 1\leq i\leq n, \label{150817:1715}
\end{equation}
is not a $(n+1)$-point function, but only a lower dimensional
$n$-point function of $(n+1)$-th order with the full-ranked
transverse (solenoidal) property (see Appendix C.2)
\begin{equation}
\Big(\nabla_{\vx_1}\otimes\cdots\otimes\nabla_{\vx_i}\otimes
\cdots\otimes\nabla_{\vx_n}\otimes\nabla_{\vx_i}\Big)\cdot
\widehat{\vH}^\perp_{i,n+1}(\vx_1,\dotsc,\vx_i,\dotsc\vx_n,\vx_i,t)=0,
\quad 1\leq i\leq n.
\end{equation}
The MPC equations \eqref{150817:1647} go along with the following
natural incompressibility constraints
\begin{equation}
\left. \begin{aligned}
\nabla_{\vx_k}\cdot\vH_n(\vx_1,\dotsc,\vx_n,t)=\boldsymbol{0},
\;\;\:\text{$\forall k$ between $1\leq k\leq n$},\text{\hspace{2.25cm}}\\[0.5em]
\nabla_{\vx_k}\cdot
\widehat{\vH}^\perp_{i,n+1}(\vx_1,\dotsc,\vx_l,\dotsc\vx_n,\vx_l,t)=\boldsymbol{0},
\;\;\:\text{$\forall k,l$ between $1\leq (k,l)\leq n$, for $k\neq
l$,} \label{150817:1650}
\end{aligned}
~~~\right\}
\end{equation}

\noindent due to the incompressibility constraint
$\nabla_{\vx_k}\cdot\vu(\vx_k,t)=0$ of the instantaneous velocity
field $\vu(\vx_k,t)$, when evaluated at each point $\vx=\vx_k$
(for all $k=1,\dotsc, n$) within the single physical domain~$\vx$.
Note that, when decomposing the transverse fields
\eqref{150817:1715} into their full fields (see Appendix C.2), the
system \eqref{150817:1647}-\eqref{150817:1650} matches the
infinite Friedmann-Keller hierarchy of MPC equations as derived in
\cite{Oberlack10} for the instantaneous case.

Now, if we collect the key properties of the combined system
\eqref{150809:1945} and \eqref{150817:1647}, the most important
thing about it is that it is linear. Both the higher level Hopf
equation \eqref{150809:1945} as well as its induced lower level
system of MPC equations \eqref{150817:1647} constitute a linear
system of equations. The former as a singly closed functional
equation, while the latter as an infinite hierarchy of PDEs. As we
will see during this study, a crucial aspect of the considered
hierarchy \eqref{150817:1647} is that it is defined forward
recursively. To illustrate the basic problems from which such an
infinite system suffers, we will investigate the following
lower-dimensional analogous system for a spatially one-dimensional
field $u=u(x,t)$ with its induced moments $u_n=u_n(t)$
\begin{gather}
\partial_t u = \partial_x^2 u-\lambda\cdot x^2\, u,\label{150817:2056}\\[0.5em]
\frac{d u_n}{dt}=n\cdot (n-1)\cdot u_{n-2}-\lambda\cdot
u_{n+2},\;\;\text{where}\;\; u_n(t)=\int_{-\infty}^\infty x^n\cdot
u(x,t)\, dx,\;\, n\geq 0,\label{150817:2057}
\end{gather}

\noindent which features all key properties of the original
combined system: The linear and non-auto\-no\-mous property of the
higher level functional Hopf equation \eqref{150809:1945} is
featured by the higher level PDE \eqref{150817:2056}, from which
the lower level ordinary differential (ODE) moment equations
\eqref{150817:2057} are induced, which themselves again feature
the linear, autonomous and forward recursive property of the
infinite Friedmann-Keller hierarchy of MPC equations
\eqref{150817:1647} which again emerge from the higher level Hopf
equation\footnote[2]{Note that the incompressibility constraints
\eqref{150817:1650} cannot be mapped down to a one-dimensional
system. Hence, the incompressibility property of the original MPC
system of equations \eqref{150817:1647} is not featured by its
analogous infinite system \eqref{150817:2057}. But this is no
drawback for the issues to be addressed in this study, since the
incompressibility constraints are irrelevant and thus do not
influence the conclusions made herein.} (see Appendix D for the
derivation of \eqref{150817:2057} from \eqref{150817:2056}).

When considering in particular the original system of moments
\eqref{150817:1647}, the foremost aim, of course, is to find
certain solutions, at least some asymptotic solutions, e.g. as
attempted by Oberlack et~al.~in applying a Lie group based
symmetry group analysis directly on system \eqref{150817:1647} in
order to generate certain specific invariant solutions which then
should function as first principle scaling laws within some
asymptotic flow regime
\citep{Oberlack10,Oberlack13.1,Oberlack14,Oberlack14.1,Oberlack14.2}.

Our claim is that system \eqref{150817:1647}, along with its
incompressibility constraints \eqref{150817:1650}, is unclosed,
even if all infinite equations are formally considered. Hence such
an approach as performed by Oberlack et al. is misleading when a
symmetry analysis is only carried out for the statistical system
of moments without additionally incorporating the underlying
deterministic set of equations. In other words, to generate any
kind of solutions directly from \eqref{150817:1647} is not well
defined if no prior modelling assumptions in accord with the
deterministic Navier-Stokes equations are made to close the
statistical system of moments.

The reason why system \eqref{150817:1647} is unclosed, even if all
infinite equations are considered, is basically twofold: The first
issue lies within the non-symmetric and thus incomplete coupling
between the equations. For each order, the unknown moment
$\widehat{\vH}_{i,n+1}^\perp$ is not directly coupled to the next
higher order equation. Although all components of that moment can
be uniquely constructed from the higher dimensional moment
$\vH_{n+1}$ once they are known, which is formally denoted as
\eqref{150817:1715}, the necessary inverse construction, however,
fails. Hence, since \eqref{150817:1715} is a non-invertible
construction, i.e. since $\vH_{n+1}$ cannot be uniquely
constructed from $\widehat{\vH}_{i,n+1}^\perp$, these latter
moments are to be identified as unclosed functions in system
\eqref{150817:1647} as they do not directly enter the next higher
order correlation equation. For each order $n$ in the hierarchy
the total number of dynamical equations \eqref{150817:1647}, along
with the continuity constraint equations \eqref{150817:1650}, is
therefore always less than the total number of unknown functions.
In total this just reflects the classical closure problem of
turbulence for the moments which cannot be bypassed by simply
establishing the formal connection  \eqref{150817:1715} --- for a
more detailed discussion on this issue, please refer to
Appendix~A~\&~C in \cite{Frewer14.1} and Section 3 in
\cite{Frewer14.3}.

The second and more fundamental issue of why the infinite
hierarchy \eqref{150817:1647} is unclosed, and which forms the
focus of this study, is that the hierarchy is defined {\it
forward} recursively. As I will elucidate at the example of the
lower-dimensional analogue \eqref{150817:2057} to system
\eqref{150817:1647}, an infinite forward-recursively defined
hierarchy of equations always leads to a non-uniqueness problem in
its solution manifold --- even when sufficient initial conditions
are posed, a system such as \eqref{150817:1647}, when stated as a
well-posed initial value problem, still does not return a unique
solution. However, this problem does {\it not} arise if any
infinite hierarchy is considered which is defined {\it backward}
recursively. Hence, instead of the full system
\eqref{150817:2056}-\eqref{150817:2057}, I will first investigate
the solution properties of the subsystem $\lambda=0$, which will
be done in the next section, thus leading us to an infinite
hierarchy of moment equations which first will be defined backward
recursively.

\section{Example for an infinite backward differential recurrence
relation\\  {\normalsize\textbf{\emph{--- A closed system with a
unique general solution manifold}}}}

In this section we will analyze the combined system
\eqref{150817:2056}-\eqref{150817:2057} for $\lambda=0$, which
hence results to a study of the usual one-dimensional diffusion
equations and its moments. Consider first the higher level PDE
\eqref{150817:2056} by stating it as the following well-posed
Cauchy problem
\begin{equation}
\partial_t u =\partial_x^2 u,\;\;\,\text{with}\;\: u(x,0)=\phi(x),
\label{150818:1155}
\end{equation}
where we assume that the function $u=u(x,t)$ is decaying
sufficiently fast at infinity, i.e.
\begin{equation}
\lim_{x\rightarrow \pm\infty} u(x,t)=0,\quad \lim_{x\rightarrow
\pm\infty} \partial_x u(x,t)=0,\quad\forall t\geq 0,
\end{equation}
and where, for convenience, we want to assume that the initial
function $\phi$ is normalized to
\begin{equation}
\int_{-\infty}^\infty \phi(x)\, dx=1.
\end{equation}
Then the initial value problem \eqref{150818:1155} has the {\it
unique} solution (see e.g. \cite{Polyanin02})
\begin{equation}
u(x,t) = \frac{1}{\sqrt{4\pi t}}\int_{-\infty}^\infty
e^{-\frac{(x-x^\prime)^2}{4t}}\phi(x^\prime) \,
dx^\prime,\;\;\text{for}\;\; t\geq 0. \label{150818:1158}
\end{equation}
If we would choose as a initial condition, for example, the
Gaussian distribution
\begin{equation}
\phi(x)=\frac{1}{\sigma\sqrt{2\pi}}e^{-\frac{1}{2}\left(\frac{x-\mu}{\sigma}\right)^2},
\quad \sigma,\,\mu\in\mathbb{R},\;\; \sigma>0, \label{150818:1905}
\end{equation}
then the solution \eqref{150818:1158} would read
\begin{equation}
u(x,t)=\frac{1}{\sqrt{2\pi(2t+\sigma^2)}}e^{-\frac{1}{2}\frac{(x-\mu)^2}{2t+\sigma^2}},
\quad t\geq 0,
\end{equation}
with its associated moments uniquely given as (shown here only up
to third order)
\begin{align}
n=0:\quad\; & u_0(t)=\int_{-\infty}^\infty x^0\cdot u(x,t)\,
dx\, =\, 1,\label{150818:1907}\\[0.5em]
n=1:\quad\; & u_1(t)=\int_{-\infty}^\infty x^1\cdot u(x,t)\, dx\,
=\,\mu,\\[0.5em]
n=2:\quad\; & u_2(t)=\int_{-\infty}^\infty x^2\cdot u(x,t)\, dx\,
=\, 2t+\mu^2+\sigma^2,\\[0.5em]
n=3:\quad\; & u_3(t)=\int_{-\infty}^\infty x^3\cdot u(x,t)\, dx\,
=\,\mu\left(6t+\mu^2+3\sigma^2\right).\label{150818:1909}
\end{align}

\noindent Now, let's consider the corresponding moment equations
when they are independently derived as an infinitely coupled
hierarchy from the higher-level Cauchy problem \eqref{150818:1155}
(see Appendix~D)
\begin{equation}
\frac{d u_n}{dt}=n\cdot (n-1)\cdot u_{n-2},\;\;\,\text{with}\;\:
u_n(0)=\int_{-\infty}^\infty x^n\cdot \phi(x)\, dx,\quad n\geq 0.
\label{150818:1441}
\end{equation}
Our aim is to investigate in how far the above infinite system of
first order ODEs represents a well-posed initial value-problem for
the moments $u_n=u_n(t)$. The first positive observation is that
to each unknown function $u_n$ one can bijectively associate an
initial condition $u_n(0)$ to it, since any first order ODE only
requires a single initial condition in order to provide a unique
solution. The next step is to derive the general solution, for
which it is helpful to recognize that since the infinite hierarchy
in \eqref{150818:1441} is defined {\it backward recursively of an
order two}, it can be rewritten into the following equivalent form
\begin{equation}
\left .
\begin{aligned}
& \frac{du_0(t)}{dt}=0;\quad \;\;\;\;\frac{d^{k+1}
u_{2k+2}(t)}{dt^{k+1}}
=(2k+2)!\cdot u_0(t),\;\; k\geq 0,\\[0.5em]
&\frac{du_1(t)}{dt}=0;\quad\;\;\;\;\frac{d^l
u_{2l+1}(t)}{dt^l}=(2l+1)!\cdot u_1(t),\;\; l\geq 1.
\end{aligned}
~~~~ \right\} \label{150818:1459}
\end{equation}

\noindent This system can then be uniquely integrated to give the
{\it general} solution
\begin{equation}
\!\!\!\!\!\left .
\begin{aligned}
u_0(t)& \, =\, c_0,\\[0.5em]
u_{2k+2}(t)
&\, = \, c_{2k+2}\, +\,\sum_{j=1}^{k}\frac{q^{(1)}_{k,j}}{j!}\, t^j\\
&\text{\hspace{1.52cm}}\, +\,
(2k+2)!\int_0^{t_k=t}\int_0^{t_{k-1}}\!\cdots\int_0^{t_0}u_0(t^\prime)\,
dt^\prime\, dt_0\cdots
dt_{k-1},\;\; k\geq 0,\\[0.5em]
\quad u_1(t)&\, =\, c_1,\\[0.5em]
u_{2l+1}(t)&\, =\, c_{2l+1}\, +\,
\sum_{j=1}^{l-1}\frac{q^{(2)}_{l,j}}{j!}\, t^j\\
&\text{\hspace{1.48cm}}\,
+\,(2l+1)!\int_0^{t_l=t}\int_0^{t_{l-1}}\!\cdots\int_0^{t_1}u_1(t^\prime)\,
dt^\prime\, dt_1\cdots dt_{l-1},\;\; l\geq 1,
\end{aligned}
~~~\right\}\label{150818:1836}
\end{equation}

\noindent with the expansion coefficients given as
\begin{equation}
\left .
\begin{aligned}
q^{(1)}_{k,j}&=\frac{(2k+2)!}{(2k+2-2j)!}\, c_{2k+2-2j},\;\; k\geq
0;\;\; 0\leq j\leq k,\\[0.5em]
q^{(2)}_{l,j}&=\frac{(2l+1)!}{(2l+1-2j)!}\, c_{2l+1-2j},\;\; l\geq
1;\;\; 0\leq j\leq l-1,
\end{aligned}
~~~ \right\}
\end{equation}

\noindent where all $c_n$ for $n\geq 0$ are arbitrary integration
constants. Hence, the second positive observation is that the {\it
unrestricted} system of equations in \eqref{150818:1441}
\begin{equation}
\frac{d u_n(t)}{dt}=n\cdot (n-1)\cdot u_{n-2}(t),\quad n\geq
0,\label{150819:1410}
\end{equation}
provides a general solution \eqref{150818:1836} which only
involves arbitrary constants, i.e. the unrestricted system
\eqref{150819:1410} provides a {\it unique} general solution and
therefore it ultimately represents a formally closed system of
equations. Because, when restricting this system to the underlying
PDE's initial condition $u(x,0)=\phi(x)$ as given in
\eqref{150818:1155}, which for the infinite ODE system takes the
integral form $u_n(0)$ as given in \eqref{150818:1441}, will turn
the general solution \eqref{150818:1836} into a unique and fully
determined solution, where the integration constants are then
given by
\begin{equation}
c_n=u_n(0), \;\; n\geq 0. \label{150820:2219}
\end{equation}
By choosing for example again the specific initial condition
\eqref{150818:1905}, the general solution \eqref{150818:1836} will
give exactly the expressions
\eqref{150818:1907}-\eqref{150818:1909} for the moments, which, as
we know, were directly derived from the higher-level PDE solution
\eqref{150818:1158} of the well-posed Cauchy problem
\eqref{150818:1155}. Hence, since the {\it unrestricted} infinite
system of ODEs \eqref{150819:1410} constitutes a formally closed
system of equations, the {\it restricted} system
\eqref{150818:1441} expresses a well-posed initial value problem.

That \eqref{150819:1410} constitutes a formally closed system is
also supported when performing a Lie-group based symmetry analysis
(see e.g. \cite{Stephani89,Olver93,Bluman96,Ibragimov94,Hydon00}).
The infinite and coupled system \eqref{150819:1410} admits the
following complete and unique set of Lie-point
symmetries\footnote[2]{The explicit result for the symmetries
\eqref{150819:1442} was obtained by augmenting the considered
system \eqref{150819:1410} to an extended system which also
includes all admissible differential consequences of
\eqref{150819:1410}. Otherwise one runs the high risk of
performing an inconsistent symmetry analysis. Note that the
symmetry corresponding to the linear superposition principle has
not been included in \eqref{150819:1442}, since it cannot be taken
to directly construct group invariant solutions as we intend to do
here.}
\begin{equation}
\left. \begin{aligned} \mathsf{S}_1: & \;\;\;
X_1=\partial_t,\\[0.5em]
\mathsf{S}_2: & \;\;\;
X_2=t\partial_t-2u_0\partial_{u_0}-u_2\partial_{u_2}
+\cdots + (n-2)\cdot u_{2n}\partial_{u_{2n}}+\cdots\\
&\text{\hspace{1.79cm}} -2u_1\partial_{u_1}-u_3\partial_{u_3}
+\cdots
+ (n-2)\cdot u_{2n+1}\partial_{u_{2n+1}}+\cdots ,\\[0.5em]
\mathsf{S}_3: & \;\;\; X_3=u_0\partial_{u_0}+u_2\partial_{u_2}
+\cdots + u_{2n}\partial_{u_{2n}}+\cdots,\\[0.5em]
\mathsf{S}_4: & \;\;\; X_4=u_1\partial_{u_1}+u_3\partial_{u_3}
+\cdots + u_{2n+1}\partial_{u_{2n+1}}+\cdots,\\[0.5em]
\mathsf{S}_5: & \;\;\;
X_5=u_1\partial_{u_0}+\frac{1}{3}u_3\partial_{u_2}+\cdots+
\frac{1}{2n+1}u_{2n+1}\partial_{u_{2n}}+\cdots,\\[0.5em]
\mathsf{S}_6: & \;\;\;
X_6=u_0\partial_{u_1}+3u_2\partial_{u_3}+\cdots+ (2n+1)\cdot
u_{2n}\partial_{u_{2n+1}}+\cdots,\\[0.5em]
\mathsf{S}^\infty_{2n}: & \;\;\;
X^\infty_{2n}=\partial_{u_{2n}}+\frac{(2n+2)!}{(2n)!}\,t\,\partial_{u_{2n+2}}
+\frac{(2n+4)!}{2!(2n)!}\,t^2\,\partial_{u_{2n+4}}+\cdots\\
&\text{\hspace{4.95cm}} +\cdots
+\frac{(2n+2m)!}{m!(2n)!}\,t^m\,\partial_{u_{2n+2m}}+\cdots,\;\:
m\geq 0,\\[0.5em]
\!\!\!\!\!\mathsf{S}^\infty_{2n+1}: & \;\;\;
X^\infty_{2n+1}=\partial_{u_{2n+1}}+\frac{(2n+3)!}{(2n+1)!}\,t\,\partial_{u_{2n+3}}
+\frac{(2n+5)!}{2!(2n+1)!}\,t^2\,\partial_{u_{2n+5}}+\cdots\\
&\text{\hspace{4.0cm}} +\cdots
+\frac{(2n+2m+1)!}{m!(2n+1)!}\,t^m\,\partial_{u_{2n+2m+1}}+\cdots,\;\:
m\geq 0,
\end{aligned}
~~~ \right \} \label{150819:1442}
\end{equation}

\noindent where $X_k$ in each case is the infinitesimal generator
of the symmetry $\mathsf{S}_k$. These generators form a {\it
closed} Lie algebra as shown in the commutator table below.
%, which serves as a further and independent indication that the infinite
%hierarchy of equations \eqref{150819:1410} can be identified as a
%closed system.
%
\begin{table}[h]
\centering
\renewcommand{\arraystretch}{2.0}
\begin{tabular}{|c|c c c c c c c c|}\hline
$[\,\cdot\, ,\cdot\,]$ & $X_1$ & $X_2$ & $X_3$ & $X_4$ & $X_5$ &
$X_6$ & $X^\infty_{2n}$ & $X^\infty_{2n+1}$
\\\hline
$X_1$ & $0$ & $X_1$ & $0$ & $0$ & $0$ & $0$ &
$\frac{(2n+2)!}{(2n)!} X^\infty_{2n+2}$ & $\frac{(2n+3)!}{(2n+1)!}
X^\infty_{2n+3}$
\\
$X_2$ & $$ & $0$ & $0$ & $0$ & $0$ & $0$ &
$\raisebox{1.0pt}{$\scriptstyle -(n-2)$} X^\infty_{2n}$ &
$\raisebox{1.0pt}{$\scriptstyle -(n-2)$} X^\infty_{2n+1}$
\\
$X_3$ & $$ & $$ & $0$ & $0$ & $\raisebox{1.5pt}{$\scriptstyle
-$}X_5$ & $X_6$ & $\raisebox{1.5pt}{$\scriptstyle
-$}X^\infty_{2n}$ & $0$
\\
$X_4$ & $$ & $$ & $$ & $0$ & $X_5$ &
$\raisebox{1.5pt}{$\scriptstyle -$}X_6$ & $0$ &
$\raisebox{1.5pt}{$\scriptstyle -$}X^\infty_{2n+1}$
\\
$X_5$ & $$ & $$ & $$ & $$ & $0$ &
$X_3\raisebox{1.25pt}{$\scriptstyle \,\, +\,\,$}X_4$ & $0$ &
$\raisebox{0.75pt}{$\scriptstyle -\,$}\frac{1}{2n+1}X^\infty_{2n}$
\\
$X_6$ & $$ & $$ & $$ & $$ & $$ & $0$ &
$\raisebox{1.0pt}{$\scriptstyle -(2n+1)$}X^\infty_{2n+1}$ & $0$
\\
$X^\infty_{2n}$ & $$ & $$ & $$ & $$ & $$ & $$ & $0$ & $0$
\\
$X^\infty_{2n+1}$ & $$ & $$ & $$ & $$ & $$ & $$ & $$ & $0$
\\
\hline
\end{tabular}
\renewcommand{\arraystretch}{1.0}
\caption{{\small Commutator table for the generators
\eqref{150819:1442}, where
$[X_i,X_j]=-[X_j,X_i]:=X_iX_j-X_jX_i$.}}
\end{table}
\noindent Since a Lie algebra inherently defines a linear vector
space, any linear combination of generators \eqref{150819:1442}
forms again a symmetry generator of the considered system
\eqref{150819:1410}. And, due to the large variety of symmetries
admitted, it is possible to combine these symmetry generators
\eqref{150819:1442} into several independent symmetries such that
they are compatible with the posed initial condition given in
\eqref{150818:1441}. Altogether three such compatible symmetries
can be constructed from \eqref{150819:1442}:
\begin{equation}
\left. \begin{aligned} \mathsf{S}^\phi_1: & \;\;\;
X_1^\phi=X_2-\sum_{k=0}^\infty (k-2)\bigg(u_{2k}(0)\cdot
X^\infty_{2k}+u_{2k+1}(0)\cdot X^\infty_{2k+1}\bigg),\\[0.5em]
\mathsf{S}^\phi_2: & \;\;\; X_2^\phi=X_3+X_4 - \sum_{k=0}^\infty
\bigg(u_{2k}(0)\cdot X^\infty_{2k}+u_{2k+1}(0)\cdot
X^\infty_{2k+1}\bigg),\\[0.5em]
\mathsf{S}^\phi_3: & \;\;\;  X_3^\phi=X_5+X_6 - \sum_{k=0}^\infty
\left(\frac{1}{2k+1}\cdot u_{2k+1}(0)\cdot
X^\infty_{2k}+(2k+1)\cdot u_{2k}(0)\cdot X^\infty_{2k+1}\right).
\end{aligned}
~~~ \right \} \label{150820:1631}
\end{equation}

\noindent When generating invariant solutions from
\eqref{150820:1631}\footnote[3]{Note that the determination of an
invariant function from any of the three symmetries
\eqref{150820:1631} is not sufficient to be a solution of the
underlying system \eqref{150818:1441}. In general such an
invariant function is endowed with integration constants which can
only be uniquely determined by plugging the function back into its
determining system of equations, i.e. here, back into system
\eqref{150818:1441}. But note that if one of these integration
constants can not be equated consistently, then the invariant
function does not constitute a solution.}, all three symmetries
independently will finally give the same {\it unique} solution set
\eqref{150818:1836}, with $c_n=u_n(0)$ \eqref{150820:2219}, as
when solving directly the system of equations \eqref{150818:1441}.

\section{Example for an infinite forward differential recurrence
relation\\  {\normalsize\textbf{\emph{--- An unclosed system with
a non-unique general solution manifold}}}}

In this section we will investigate an infinite system which is
not closed; which yet not even formally can be regarded as closed.
By analyzing the combined system
\eqref{150817:2056}-\eqref{150817:2057} for $\lambda=1$, we are
now obtaining a one-dimensional system which, in contrast to the
previously in Section~3 considered system, now features {\it all}
key properties of the higher level induced (Hopf-equation induced)
infinite Friedmann-Keller hierarchy of MPC
equations~\eqref{150817:1647}, in particular where the crucial
property of being a forward recursively defined relation is now
included. And it is solely this property which now turns the
problem into an unclosed~problem.

As it was done in the previous section for the specification
$\lambda=0$, let us also for $\lambda=1$ first consider the higher
level PDE \eqref{150817:2056} as the following well-posed Cauchy
problem
\begin{equation}
\partial_t u=\partial_x^2 u-x^2 u,\;\:\text{for}\;\: t\geq
0,\;\;\text{with}\;\; u(x,0)=\phi(x),\label{150821:1049}
\end{equation}
where we again assume that the function $u=u(x,t)$ is decaying
sufficiently fast at infinity, i.e.
\begin{equation}
\lim_{x\rightarrow \pm\infty} u(x,t)=0,\quad \lim_{x\rightarrow
\pm\infty} \partial_x u(x,t)=0,\quad\forall t\geq 0,
\end{equation}
and where also again, for convenience, we want to assume that the
initial function $\phi$ is nor\-malized to one
\begin{equation}
\int_{-\infty}^\infty \phi(x)\, dx=1.
\end{equation}
To solve this initial value problem \eqref{150821:1049} it is
necessary to realize that the following nonlinear point
transformation \citep{Polyanin02}\footnote[2]{This continuous
point transformation is {\it not} a group transformation, as it
neither includes a group parameter nor does it include the unique
continuously connected identity transformation from which any
infinitesimal mapping can emanate.}
\begin{equation}
\tilde{t}=\frac{1}{4}\cdot\left(e^{4t}-1\right),\quad
\tilde{x}=x\cdot e^{2t},\quad \tilde{u}=u\cdot
e^{-\frac{1}{2}x^2-t}, \label{150821:1113}
\end{equation}
which has the unique inverse transformation
\begin{equation}
t=\frac{1}{4}\ln\big(1+4\tilde{t}\hspace{.06cm}\big),\quad
x=\frac{\tilde{x}}{\sqrt{1^{\vphantom{A^A}}+4\tilde{t}\,}},\quad
u=\tilde{u}\cdot \sqrt[4]{1+4\tilde{t}\,}\cdot
e^{\frac{1}{2}\cdot\frac{\tilde{x}^2}{1^{\vphantom{1}}+4\tilde{t}}},
\end{equation}
maps the original Cauchy problem \eqref{150821:1049} into the
following Cauchy problem for the standard diffusion equation with
constant coefficients:
\begin{equation}
\partial_{\tilde{t}\,}
\tilde{u}=\partial_{\tilde{x}\,}^2\tilde{u},\;\:\text{for}\;\:
\tilde{t}\geq 0,\;\;\text{with}\;\;
\tilde{u}(\tilde{x},0)=\phi(\tilde{x})\cdot
e^{-\frac{1}{2}\tilde{x}^2}. \label{150821:1114}
\end{equation}
Important to note here is that the initial time $t=0$ as well as
the relevant time range $t\in [0,\infty)$ both get invariantly
mapped to $\tilde{t}=0$ and $\tilde{t}\in [0,\infty)$
respectively. Hence, the unique solution of the transformed Cauchy
problem \eqref{150821:1114} is thus again given by
\eqref{150818:1158}, but now in the form
\begin{equation}
\tilde{u}(\tilde{x},\tilde{t}) =
\frac{1}{\sqrt{4\pi^{\vphantom{A}}
\tilde{t}\,}}\int_{-\infty}^\infty
e^{-\frac{(\tilde{x}-\tilde{x}^\prime)^2}{4^{\vphantom{1}}\tilde{t}}}\phi(\tilde{x}^\prime)
\, e^{-\frac{1}{2}\tilde{x}^{\prime\, 2}} \,
d\tilde{x}^\prime,\;\;\text{for}\;\; \tilde{t}\geq 0,
\label{150821:1121}
\end{equation}
which then, according to transformation \eqref{150821:1113}, leads
to the {\it unique} solution for the original Cauchy
problem~\eqref{150821:1049}
\begin{equation}
u(x,t)=
\frac{e^{\frac{1}{2}x^2+t}}{\sqrt{\pi\big(e^{4t}-1^{\vphantom{1}}\big)}}
\int_{-\infty}^\infty
e^{-\frac{(e^{2t}x-x^\prime)^2}{e^{4t^{\vphantom{:}}}-1}}\phi(x^\prime)
\,\, e^{-\frac{1}{2}x^{\prime\, 2}} \,
dx^\prime,\;\;\text{for}\;\; t\geq 0. \label{150821:1122}
\end{equation}
If we again would choose as a initial condition, for example, the
Gaussian distribution \eqref{150818:1905}
\begin{equation}
\phi(x)=\frac{1}{\sigma\sqrt{2\pi}}e^{-\frac{1}{2}\left(\frac{x-\mu}{\sigma}\right)^2},
\quad \sigma,\,\mu\in\mathbb{R},\;\; \sigma>0, \label{150821:1155}
\end{equation}
then the solution \eqref{150821:1122} would have the explicit form
\begin{equation}
u(x,t)=\frac{1}{\sqrt{\pi\big(\sigma^2+e^{4t}(1+\sigma^2)-1\big)}}
e^{t+\frac{x^2}{2}-\frac{e^{4t}(\mu^2+2x^2(1+\sigma^2))-4e^{2t}\mu
x+\mu^2}{2(\sigma^2+e^{4t}(1+\sigma^2)-1)}}, \quad t\geq 0.
\label{150821:1956}
\end{equation}
Its moments for {\it all} $t\geq 0$ are uniquely given as (shown
here again only up to third order)
\begin{align}
n=0:\quad\; & u_0(t)=\int_{-\infty}^\infty x^0\cdot u(x,t)\, dx\,
=\, \frac{\sqrt{2}\, e^{t-
\frac{(-1+e^{4t})\mu^2}{2(1-\sigma^2+e^{4t}(1+\sigma^2))}}}
{\sqrt{1-\sigma^2+e^{4t}\big(1+\sigma^2\big)}},\label{150821:1711}\\[0.5em]
n=1:\quad\; & u_1(t)=\int_{-\infty}^\infty x^1\cdot u(x,t)\, dx\,
=\,\frac{2\mu\sqrt{2}\, e^{3t-
\frac{(-1+e^{4t})\mu^2}{2(1-\sigma^2+e^{4t}(1+\sigma^2))}}}
{\sqrt{\big(1-\sigma^2+e^{4t}\big(1+\sigma^2\big)\big)^3}},\\[0.5em]
n=2:\quad\; & u_2(t)=\int_{-\infty}^\infty x^2\cdot u(x,t)\, dx\,
=\,  \frac{\sqrt{2}\, e^{t-
\frac{(-1+e^{4t})\mu^2}{2(1-\sigma^2+e^{4t}(1+\sigma^2))}}}
{\sqrt{\big(1-\sigma^2+e^{4t}\big(1+\sigma^2\big)\big)^5}}\nonumber\\
&\text{\hspace{5cm}}\cdot
\Big(4e^{4t}\mu^2-(-1+\sigma^2)^2+e^{8t}(1+\sigma^2)^2\Big)
,\\[0.5em]
n=3:\quad\; & u_3(t)=\int_{-\infty}^\infty x^3\cdot u(x,t)\, dx\,
=\,\frac{2\mu\sqrt{2}\, e^{3t-
\frac{(-1+e^{4t})\mu^2}{2(1-\sigma^2+e^{4t}(1+\sigma^2))}}}
{\sqrt{\big(1-\sigma^2+e^{4t}\big(1+\sigma^2\big)\big)^7}}\nonumber\\
&\text{\hspace{5cm}}\cdot
\Big(4e^{4t}\mu^2-3(-1+\sigma^2)^2+3e^{8t}(1+\sigma^2)^2\Big).\label{150821:1713}
\end{align}

\noindent Now, let's consider the corresponding moment equations
when they are independently derived from the higher-level Cauchy
problem \eqref{150821:1049} (see Appendix~D)
\begin{equation}
\frac{du_n}{dt} = n\cdot (n-1)\cdot
u_{n-2}-u_{n+2},\;\;\,\text{with}\;\: u_n(0)=\int_{-\infty}^\infty
x^n\cdot \phi(x)\, dx,\quad n\geq 0. \label{150821:1818}
\end{equation}
Our aim is again to investigate in how far the above infinite
system of first order ODEs represents a well-posed initial
value-problem for the moments $u_n=u_n(t)$. As it was also already
noticed for the system in the previous section, the first positive
observation is that in this case too, one can bijectively
associate an initial condition $u_n(0)$ to each unknown function
$u_n$. But the next step, deriving the general solution of
\eqref{150821:1818}, stands in clear contrast to the previous
section: The system's general solution is {\it not} uniquely
specified, i.e.~even if sufficient initial conditions are imposed,
system \eqref{150821:1818} still does not offer a unique solution.
In clear contrast, of course, to its associated higher level PDE
system \eqref{150821:1049}, which, as a Cauchy problem, is
well-posed by providing the unique solution \eqref{150821:1956}
with its unique moments \eqref{150821:1711}-\eqref{150821:1713}.
To see this problem explicitly, it is helpful to first recognize
that the infinite ODE system \eqref{150821:1818} can be rewritten
into the following equivalent and already solved
form\footnote[2]{In the following we agree on the definitions that
$\frac{d^0}{dt^{0}}=1$, $\frac{d^{q<0}}{dt^{q}}=0$, and
$\sum_{i=0}^{q<0}=0$.}
\begin{equation}
\left.
\begin{aligned}
u_{2k+2}(t)&=(-1)^{k+1}\sum_{i=0}^\infty
A^{(1)}_i(k)\frac{d^{\hspace{0.5pt}k+1-2i}}{dt^{\hspace{0.2pt}k+1-2i}}\,
u_0(t),\;\; k\geq
0,\\[0.5em]
u_{2l+1}(t)&=(-1)^{l}\sum_{j=1}^\infty
A^{(2)}_j(l)\frac{d^{\hspace{0.75pt}
l+2-2j}}{dt^{\hspace{0.2pt}l+2-2j}}\, u_1(t),\;\; l\geq 1,
\end{aligned}
~~~ \right\}\label{150821:1859}
\end{equation}

\noindent where the coefficients $A^{(1)}_i(k)$ and $A^{(2)}_j(l)$
are recursively defined as:
\begin{equation}
\left.
\begin{aligned}
A^{(1)}_i(k) &= \sum_{q=0}^{k-(2i-1)} (2k-2q)\cdot (2k-1-2q)\cdot
A^{(1)}_{i-1}(k-2-q),\;\;\; i\geq 1,\;\; k\geq 0,
\\[0.5em]
A^{(2)}_j(l) &= \sum_{q=1}^{l-(2j-3)} (2l-2q)\cdot (2l+1-2q)\cdot
A^{(2)}_{j-1}(l-1-q),\;\;\; j\geq 2,\;\; l\geq 1,
\end{aligned}
~~~\right\}
\end{equation}

\noindent with the initial seeds: $A^{(1)}_0(-1)=1$,
$A^{(1)}_0(k)=1$, $\forall k\geq0$, and $ A^{(2)}_1(0)=1$,
$A^{(2)}_1(l)=1$, $\forall l\geq 1$, respectively. In contrast to
the integrated {\it general} solution \eqref{150818:1836} of the
previously considered {\it unrestricted} system
\eqref{150819:1410}, we see that the degree of underdeterminedness
in the above derived {\it general} solution \eqref{150821:1859} is
fundamentally different and higher than in
\eqref{150818:1836}.~Instead of integration constants $c_n$, we
now have two integration functions $u_0(t)$ and $u_1(t)$ which can
be chosen freely. Their (arbitrary) specification will then
determine all other solutions for $k\geq 0$ and $l\geq 1$
according to \eqref{150821:1859}. The reason for having two free
functions and not infinitely many free constants is that the {\it
unrestricted} system of equations in \eqref{150821:1818}
\begin{equation}
\frac{du_n(t)}{dt} = n\cdot (n-1)\cdot u_{n-2}(t)-u_{n+2}(t),\quad
n\geq 0, \label{150822:1027}
\end{equation}
defines a {\it forward} recurrence relation (of order
two)\footnote[2]{The order of the recurrence relation is defined
relative to the differential operator.} that needs {\it not} to be
integrated in order to determine its general solution, while
system \eqref{150819:1410}, in contrast, defines a {\it backward}
recurrence relation (of order two) which needs to be integrated to
yield its general solution.

To explicitly demonstrate that \eqref{150821:1859} is not a {\it
unique} general solution, we have to impose the initial condition
$u(x,0)=\phi(x)$ for the corresponding moments as given in
\eqref{150821:1818}. The easiest and most apparent way to perform
this implementation would be in a first choice to choose the two
arbitrary functions $u_0(t)$ and $u_1(t)$ as analytical functions,
since they  straightforwardly allow for an expansion into power
series
\begin{equation}
u_0(t)=\sum_{p=0}^\infty \frac{c^{(1)}_p}{p!}\, t^p,\quad\;\;
u_1(t)=\sum_{p=0}^\infty \frac{c^{(2)}_p}{p!}\, t^p,
\label{150822:1110}
\end{equation}
where $c^{(1)}_p$ and $c^{(2)}_p$ are two different infinite sets
of constant expansion coefficients. By inserting this Ansatz into
the general solution \eqref{150821:1859} and imposing the initial
conditions as given in \eqref{150821:1818} will then uniquely
specify these coefficients in a recursive manner as
\begin{equation}
\left .
\begin{aligned}
c^{(1)}_p =0,\;\; p<0; \quad\;\;\; c^{(1)}_p & =(-1)^p\cdot
u_{2p}(0)-\sum_{r=1}^\infty
A^{(1)}_r(p-1)\cdot c^{(1)}_{p-2r},\;\;p\geq 0,\\[0.5em]
c^{(2)}_p =0,\;\; p<0; \quad\;\;\; c^{(2)}_p & =(-1)^p\cdot
u_{2p+1}(0)-\sum_{r=1}^\infty A^{(2)}_{r+1}(p)\cdot
c^{(2)}_{p-2r},\;\;p\geq 0.
\end{aligned}
~~~ \right\}\label{150822:1111}
\end{equation}

\noindent Indeed, the two functions \eqref{150822:1110} with the
above determined coefficients \eqref{150822:1111} form the
analytical part of the corresponding unique PDE moment solutions
\eqref{150821:1711}-\eqref{150821:1713} when choosing the
initial-condition function $\phi$ as the Gaussian distribution
\eqref{150821:1155}. In other words, if the unique PDE moment
solutions \eqref{150821:1711}-\eqref{150821:1713} were Taylor
expanded around $t=0$, they would exactly yield the first four
power series solutions \eqref{150822:1110} of the associated
infinite ODE system \eqref{150821:1818}. But, the obvious problem
is that the underlying PDE solutions
\eqref{150821:1711}-\eqref{150821:1713} do {\it not} constitute
analytical functions in the independent variable $t$, i.e.~the
Taylor expansions of the unique solution functions
\eqref{150821:1711}-\eqref{150821:1713} do not converge on the
whole temporal domain, but only in the following limited range
(dependent, of course, on the value of the variance $\sigma^2$)
\begin{equation}
\left.
\begin{aligned}
 0<\sigma^2<1\! : &\qquad 0\leq t < \frac{1}{4}\sqrt{\pi^{2^{\vphantom{.}}}
+\left[\ln\left(\frac{1-\sigma^2}{1+\sigma^2}\right) \right]^2
\,},\\[0.5em]
\sigma^2=1\! : &\qquad 0\leq t<\infty,\\[0.5em]
\sigma^2>1\! : &\qquad 0\leq t <
\frac{1}{4}\sqrt{\big(2\pi\big)^{2^{\vphantom{.}}}
+\left[\ln\left(\frac{\sigma^2-1}{\sigma^2+1}\right) \right]^2
\,}.
\end{aligned}
~~~~ \right \}
\end{equation}

\noindent That means, our initial assumption that the first two
ODE solutions $u_0(t)$ and $u_1(t)$ are\linebreak analytical
functions on the global and unlimited scale $t\in[0,\infty)$, for
all values of $\sigma^2$, is thus not correct. Only for a very
limited range, where $\sigma^2\sim 1$, this functional choice
\eqref{150822:1110} is valid. But, if we don't know the full scale
PDE solutions \eqref{150821:1711}-\eqref{150821:1713} beforehand,
how then to choose these two unknown functions $u_0(t)$ and
$u_1(t)$ for the infinite ODE system \eqref{150821:1859}? The
clear answer is that there is no way without invoking a prior
modelling assumption on the ODE system itself. Even if we would
choose specific functions $f_0(t)$ and $f_1(t)$, which for
$u_0(t)$ and $u_1(t)$ are valid on any larger scale than the
limited analytical Ansatz \eqref{150822:1110}, we still have the
problem that this particular solution choice is not unique,
because one can always add to this choice certain independent
functions which give no contributions when evaluated at the
initial point $t=0$. For example, if $u_0(t)=f_0(t)$ and
$u_1(t)=f_1(t)$, and if both functions $f_0$ and $f_1$ satisfy the
given initial conditions at $t=0$,~then
\begin{equation}
u_0(t)=f_0(t)+ \psi_0(t)\cdot e^{-\frac{\gamma_0^2}{t^2}},\qquad
u_1(t)=f_1(t)+ \psi_1(t)\cdot e^{-\frac{\gamma_1^2}{t^2}},
\label{150823:1657}
\end{equation}
is also a possible solution choice which satisfies the same
initial conditions, where $\psi_0(t)$ and~$\psi_1(t)$ are again
(new) arbitrary functions, with the only restriction that, at the
initial point $t=0$, they have to increase slower than
$e^{\gamma_0^2/t^2}$ and $e^{\gamma_1^2/t^2}$ respectively.

Note that $u_0(t)$ and $u_1(t)$ are not privileged in the sense
that only these functions can be chosen arbitrarily. Any two
functions $u_{n^*}(t)$ and $u_{m^*}(t)$ in the hierarchy
\eqref{150822:1027} can be chosen freely, where $n=n^*$ and
$n=m^*$ are some arbitrary but fixed orders in this hierarchy such
that $n^*\neq m^*+2k$, $\forall k\in\mathbb{Z}$. For example, for
the choice $n^*=2$ and $m^*=3$ in \eqref{150822:1027} its
underdetermined general solution will take the form
\begin{equation}
\left.
\begin{aligned}
u_0(t)&=u_0(0)-\int_0^t u_2(t^\prime)dt^\prime,\\[0.5em]
u_{2k+2}(t)&=(-1)^{k+1}\sum_{i=0}^\infty
A^{(1)}_i(k)\left[\delta_{0,k+1-2i}\left(u_0(0)-\int_0^t
u_2(t^\prime)dt^\prime\right)-\frac{d^{\hspace{0.5pt}k-2i}}{dt^{\hspace{0.2pt}k-2i}}\,
u_2(t)\right],\;\; k\geq
1,\\[0.5em]
u_1(t)&=u_1(0)-\int_0^t u_3(t^\prime)dt^\prime,\\[0.5em]
u_{2l+1}(t)&=(-1)^{l}\sum_{j=1}^\infty
A^{(2)}_j(l)\left[\delta_{0,l+2-2j}\left(u_1(0)-\int_0^t
u_3(t^\prime)dt^\prime\right)-\frac{d^{\hspace{0.75pt}
l+1-2j}}{dt^{\hspace{0.2pt}l+1-2j}}\, u_3(t)\right],\;\; l\geq 2,
\end{aligned}
~~~ \right\}\label{150822:1753}
\end{equation}

\noindent where now, compared to the alternative general solution
\eqref{150821:1859}, the functions $u_2(t)$ and $u_3(t)$ are the
unclosed terms, instead of $u_0(t)$ and $u_1(t)$.

That the infinite system \eqref{150822:1027} is underdetermined
and thus unclosed is also supported when performing an invariance
analysis on it. To transform system \eqref{150822:1027}
invariantly, complete arbitrariness exists in that three arbitrary
functions are available in order to perform the transformation:
One for the independent variable $t$ and two for any arbitrary but
fixed chosen dependent variables $u_{n^*}$ and $u_{m^*}$, such
that $n^*\neq m^*+2k$, $\forall k\in\mathbb{Z}$, i.e.~where in
effect the transformation of two functions $u_{n^*}=u_{n^*}(t)$
and $u_{m^*}=u_{m^*}(t)$ can be chosen absolutely
freely.\linebreak Note that the outcome of such an (unclosed)
invariance analysis does not result into sym\-metry
transformations, but only into weaker {\it equivalence}
transformations which invariantly only map between unclosed
systems (see also the discussion partly done in the Introduction;
for more details on this issue, see \cite{Frewer14.1,Frewer14.3}
and the references therein).

Hence, when choosing, for example, $n^*=0$ and $m^*=1$, then
infinitely many functionally independent equivalence
transformations of the following form are admitted by system
\eqref{150822:1027}
\begin{align}
\!\!\!\!\mathsf{E}^\infty_{(\xi,\eta_{u_0},\eta_{u_1})} : &\;\;\;
X\, =\, \xi\partial_t\, +\,\eta_{u_0}\partial_{u_0}
\, +\,\eta_{u_1}\partial_{u_1}\nonumber\\[0.0em]
&\hspace{1.875cm} +\,\eta_{u_2}\big(\xi,\eta_{u_0},
\eta_{u_1},u_2,u_3\big)\partial_{u_2}\nonumber\\[0.25em]
&\hspace{1.875cm} +\,\eta_{u_3}\big(\xi,\eta_{u_0},
\eta_{u_1},u_2,u_3\big)\partial_{u_3}\nonumber\\[0.25em]
&\hspace{1.875cm} +\,\cdots\, +\,
\eta_{u_{2n}}\big(\xi,\eta_{u_0}, \eta_{u_1},u_2,u_3,\dotsc
,u_{2n},u_{2n+1}\big)\partial_{u_{2n}}\nonumber\\[0.25em]
&\hspace{2.921cm} +\, \eta_{u_{2n+1}}\big(\xi,\eta_{u_0},
\eta_{u_1},u_2,u_3,\dotsc ,u_{2n},u_{2n+1}\big)\partial_{u_{2n+1}}
\, +\, \cdots ,\label{150823:1206}
\end{align}

\noindent where $\xi=\xi(t,u_0,u_1)$, $\eta_{u_0}(t,u_0,u_1)$ and
$\eta_{u_1}(t,u_0,u_1)$ are arbitrary, free choosable functions.
Consequently, the infinite set of invariant transformations
\eqref{150823:1206} do {\it not} form a closed Lie algebra (in
contrast to the infinite set \eqref{150819:1442} of the previously
considered system). In particular, when imposing the initial
condition $u(x,0)=\phi(x)$ for the corresponding moments as given
in \eqref{150821:1818}, then infinitely many and functionally
independent invariant (equivalence) transformations can be
constructed which all are compatible with this arbitrary but
specifically chosen initial condition. Because, since e.g.~the
infinitesimals $\xi=\xi(t,u_0,u_1)$, $\eta_{u_0}(t,u_0,u_1)$ and
$\eta_{u_1}(t,u_0,u_1)$ can be chosen arbitrarily, one only has to
guarantee that each of the initial\linebreak conditions
$u_n(t=0)=u_n(0)$, for all $n\geq 0$, gets invariantly mapped into
itself. This is achieved by demanding all infinitesimals to
satisfy the restrictions
\begin{equation}
\left.
\begin{aligned}
\xi(t,u_0,u_1)\Big\vert_{\{t=0;u_n=u_n(0),\forall n\geq 0\}}=0,\hspace{3cm}\\[0.5em]
\eta_{u_0}(t,u_0,u_1)\Big\vert_{\{t=0;u_n=u_n(0),\forall n\geq
0\}}=0,\qquad
\eta_{u_1}(t,u_0,u_1)\Big\vert_{\{t=0;u_n=u_n(0),\forall n\geq
0\}}=0,
\end{aligned}
~~~ \right\}\label{150823:1519}
\end{equation}
\begin{equation}
\left.
\begin{aligned}
\eta_{u_{2n}}(\xi,\eta_{u_0}, \eta_{u_1},u_2,u_3,\dotsc
,u_{2n},u_{2n+1})\Big\vert_{\{t=0;u_n=u_n(0),\forall n\geq
0\}}=0,\;\; n\geq 1,\hspace{0.625cm}\\[0.5em]
\eta_{u_{2n+1}}(\xi,\eta_{u_0}, \eta_{u_1},u_2,u_3,\dotsc
,u_{2n},u_{2n+1})\Big\vert_{\{t=0;u_n=u_n(0),\forall n\geq
0\}}=0,\;\; n\geq 1,\hspace{0.65cm}
\end{aligned}
~~~~ \right\}\!\!\!\!\!\label{150823:1520}
\end{equation}

\noindent where only the three infinitesimals $\xi$, $\eta_{u_0}$
and $\eta_{u_1}$ can be chosen freely, while the remaining
infinitesimals $\eta_{2n}$ and $\eta_{2n+1}$, for all $n\geq 1$,
are predetermined differential functions of their indicated
arguments. The conditions \eqref{150823:1519}, in accordance with
\eqref{150823:1520}, can be easily fulfilled e.g.~by restricting
the three arbitrary functions $\xi$, $\eta_{u_0}$ and $\eta_{u_1}$
to
\begin{equation}
\left.
\begin{aligned}
\xi(t,u_0,u_1)=f_0(t,u_0,u_1)\cdot
e^{-\frac{\gamma_f^2}{t^2}},\hspace{4.25cm}\\[0.5em]
\!\!\!\!\!\eta_{u_0}(t,u_0,u_1)=g_0(t,u_0,u_1)\cdot
e^{-\frac{\gamma_g^2}{(u_0-u_0(0))^2}},\;\;
\eta_{u_1}(t,u_0,u_1)=h_0(t,u_0,u_1)\cdot
e^{-\frac{\gamma_h^2}{(u_1-u_1(0))^2}},\label{150823:1530}
\end{aligned}
~ \right\}
\end{equation}

\noindent where $f_0$, $g_0$ and $h_0$ are again (new) arbitrary
functions, however, now restricted to the class of functions which
are increasing slower than $e^{1/r^2}$ at $r=0$, where
\begin{equation}
r=\sqrt{\,
t^2/\gamma_f^2+(u_0-u_0(0))^2/\gamma_g^2+(u_1-u_1(0))^2/\gamma_h^2}.
\end{equation}
And, since in this case all differential functions $\eta_{u_{2n}}$
and $\eta_{u_{2n+1}}$, $\forall n\geq 1$ have the special
non-shifted affine property
\begin{equation}
\eta_{u_{2n}}\Big\vert_{\{\xi=0;\,\eta_{u_0}=0;\,\eta_{u_1}=0\}}
=\eta_{u_{2n+1}}\Big\vert_{\{\xi=0;\,\eta_{u_0}=0;\,\eta_{u_1}=0\}}=0,\quad
\forall n\geq 1,
\end{equation}
the conditions \eqref{150823:1520} all are automatically satisfied
by the above restriction \eqref{150823:1530}. Hence, an {\it
infinite} set of functionally independent (non-privileged)
invariant solutions for each moment $u_n=u_n(t)$, $\forall n\geq
0$ can be constructed from~\eqref{150823:1530}, which all satisfy
the given initial condition~$u_n(t=0)=u_n(0)$ as imposed in
\eqref{150821:1818}.

That a priori no unique general solution, nor any unique invariant
solution can be constructed, provides the reason that the PDE
induced ODE system \eqref{150822:1027}, although infinite in
dimension, has to be treated as an unclosed system. It involves
more unknown functions than there are determining equations,
although {\it formally}, in a bijective manner, to each function
within the hierarchy a corresponding equation can be mapped to.
But, since the hierarchy \eqref{150822:1027} can be equivalently
rewritten into the solved form \eqref{150821:1859} (or into
\eqref{150822:1753}, etc.) it explicitly reveals the fact that
exactly two functions in this hierarchy, e.g. $u_0(t)$ and
$u_1(t)$, remain unknown, and without the precise knowledge of
their global functional structure all remaining solutions $u_n(t)$
for $n\geq 2$ then remain unknown too. Even when posing sufficient
initial conditions is not enough to yield a unique solution for
the lower-level ODE system \eqref{150821:1818} as it is for the
higher-level PDE equation \eqref{150821:1049}. Without a prior
modelling assumption on the ODE system \eqref{150821:1818}, this
system remains unclosed. Fortunately, the solutions of this
particular case \eqref{150821:1711}-\eqref{150821:1713} possessed
an analytical part in their functions for which the assumed Ansatz
\eqref{150822:1110} expressed the correct functional behavior,
though only in a very narrow and limited range. But, of course,
for more general cases such a partial analytical structure is not
always necessarily provided, and an Ansatz as \eqref{150822:1110}
would then be misleading.\footnote[2]{Note that not only an
analytical Ansatz as \eqref{150822:1110} would be misleading, but
also any other, functionally different Ansatz can be misleading
too, if no further external (exogenous) information is at hand.
For example, choosing instead of an analytical a Fourier-expansion
Ansatz would globally fail as the underlying functions
\eqref{150821:1711}-\eqref{150821:1713} do not show any
time-periodic behavior, at all.}

Hence, the infinite forward recursively defined system of order
two \eqref{150822:1027} does not possess a {\it unique} general
solution. The degree of arbitrariness in having two unknown
functions cannot be reduced, even when a sufficient number of
initial conditions are imposed, simply due to the existing modus
operandi \eqref{150823:1657} when constructing possible valid
solutions. The same non-uniqueness problem we face for invariant
solutions due to the construction principle \eqref{150823:1530},
in that infinitely many functionally independent equivalence
transformations can be constructed which all can be made
compatible to any specifically chosen initial condition. Hence,
the existence of an invariant solution for such a system is
without any value, since it just represents an arbitrary solution
among infinitely many other, equally privileged invariant
solutions.

\section{Conclusion}

At the example of a lower-dimensional ODE-system this study has
shown that an infinite and {\it forward} recursive hierarchy of
differential equations carries all features of an unclosed system,
and that, conclusively, all admitted invariance transformations
may only be identified as equivalence transformations. To obtain
from such systems an invariant solution which should show a
certain particular functional structure is ultimately without
value, since infinitely many functionally different and
non-privileged invariant solutions can be constructed. In order to
obtain valuable results, the infinite system needs to be closed by
posing modelling assumptions which have to reflect the structure
of the underlying (higher-level) equations from which the infinite
system~emerges.

Although in this study only an example of a lower-dimensional
system has been considered, it is well to be expected that the
very same non-uniqueness issues also occur in the
higher-dimensional Friedmann-Keller hierarchy of moments for
turbulent flows, and that it would be instructive to remember this
example when trying to generate solutions or invariant solutions
from the infinite and unmodelled Friedmann-Keller equations as its
motivated e.g.~in the work of Oberlack et al. Besides the problem
of being defined forward recursively, the infinite
Friedmann-Keller hierarchy faces another and further substantial
problem, namely the singular problem of {\it not} showing a
symmetric and thus complete coupling among all equations.~As it
was discussed at the end of Section 2, the problem that in each
equation the occurring unknown higher order moment is not directly
coupled to the corresponding next higher order equation,
additionally even increases the degree of underdeterminedness of
the Friedmann-Keller equations, which thus undoubtedly renders it
altogether to an unclosed system.

Hence, the approach and the conclusions given in
\cite{Oberlack10,Oberlack13.1,Oberlack14,Oberlack14.1,Oberlack14.2},
about deriving physically relevant invariant solutions as
so-called turbulent scaling-laws from first principles for the
lower level moments within the infinite Friedmann-Keller
hierarchy, is heavily misleading; all the more so as these first
principle invariant solutions are additionally based on physically
inconsistent symmetries (for more details, see
\cite{Frewer14.1,Frewer14.3}).\linebreak The general argument that
these turbulent scaling laws are "clearly validated" in
\cite{Oberlack11.3,Oberlack11.4,Oberlack13.1,Oberlack14,Oberlack14.2},
by comparing to direct numerical simulations (DNS), is based on a
fallacy (see Section~5 in \cite{Frewer14.1}). The problem is that
this "validation" is always only performed for the lowest order
moments, which, of course, can be robustly matched to the DNS data
since there are enough free group parameters available to be
fitted; in the very same sense as John von Neumann famously
remarked many years ago (as~quoted by Enrico Fermi to
Freeman~Dyson, see e.g. \cite{Dyson04}):

\vspace{0.5em}\noindent\emph{\phantom{x}"With four parameters I
can fit an elephant, and with five I can make him wiggle his
trunk."}

\vspace{0.5em}\noindent But, as soon as the higher order
correlations functions get fitted, the curve-fitting procedure of
Oberlack et al. consistently fails. This outcome is supported by a
mathematical proof given in \cite{Frewer14.1} (see Appendix D),
which clearly shows that the Lie-group based turbulent scaling
laws for all higher order velocity correlations as derived in
\cite{Oberlack10,Oberlack13.1,Oberlack14,Oberlack14.1,Oberlack14.2}
are {\it not} consistent to the scaling of the lowest order
correlation function, which for shear flows is the mean velocity
field. In other words, the proof in \cite{Frewer14.1} shows that
for the lowest correlation order $n = 1$ no contradiction exists,
only as from $n\geq 2$ onwards the contradiction starts, i.e.
while the mean velocity field can be robustly matched to the DNS
data, it consistently fails for all higher order correlation
functions and gets more pronounced the higher the correlation
order $n$ is.

Hence, by always only matching or comparing the DNS data to the
lowest order velocity correlation of a specific flow, as it was
done particulary in
\cite{Oberlack11.3,Oberlack11.4,Oberlack13.1,Oberlack14,Oberlack14.2},
i.e. by matching or comparing only to the mean velocity profile
for wall-bounded (shear) flows and only to the two-point velocity
correlation for isotropic flows, is definitely not enough to be a
true validation of the Lie-group based scaling theory. In
particular, as this theory by Oberlack et al. which specifically
considers the infinite (unclosed) system of all multi-point
correlation equations and which thus is designed and laid-out to
be a scaling theory for all higher order velocity correlations,
special attention has to be devoted to the prediction value of all
those correlation functions which go beyond the lowest order. And
exactly this has been investigated in \cite{Frewer14.1} (Section
5), which then gives a completely different picture than the
"validation" procedure in
\cite{Oberlack11.3,Oberlack11.4,Oberlack13.1,Oberlack14,Oberlack14.2}
is trying to suggest.

Taken together, the key reason for this failure is twofold: i)
Oberlack et al.~considers the infinite and forward recursive
Friedmann-Keller hierarchy of moments as a closed system, which,
as was clarified in this study at an analogous lower-dimensional
system of ODEs, is not the case; ii) Oberlack et al.~considers all
layers of statistical description of turbulence endogenously
without involving at the same time the underlying deterministic
Navier-Stokes system --- a methodological approach which obviously
is incomplete and even incorrect, because it are the deterministic
equations which due to their spatially nonlocal and temporally
chaotic behavior induce the statistical equations and not vice
versa. In other words, in order to obtain valuable and physically
consistent results within a statistical description of the
physically quantifiable world, it is necessary to include all
underlying processes up to the fluctuating level into this
description and not to ignore them if they exist.

\appendix

\section{Alternative form of a transverse vector field}

The transverse component $\vA^{\!\perp}(\vx)$ of an arbitrary
vector field $\vA(\vx)$, when decomposed into a longitudinal and
transverse component
$\vA(\vx)=\vA^{\!\raisebox{1.5pt}{$\scriptscriptstyle\parallel$}}(\vx)
+\vA^{\!\perp}(\vx)$ respectively, and, which decays sufficiently
fast for $|\vx|\rightarrow\infty$, is defined as (see e.g.
\cite{Stewart08})
\begin{equation}
\vA^{\!\perp}(\vx):=\vA(\vx)+\nabla\int\frac{\nabla^\prime\cdot
\vA(\vx^\prime)}{4\pi|\vx-\vx^\prime|}\, d^3\vx^\prime,
\label{150809:1528}
\end{equation}
confirming its defining (solenoidal) property
$\nabla\cdot\vA^{\!\perp}(\vx)=0$, when making use of the standard
identity
\begin{equation}
\delta^3(\vx-\vx^\prime)=-\Delta\frac{1}{4\pi|\vx-\vx^\prime|}.
\end{equation}
Relation \eqref{150809:1528} can be written in an alternative
form, by rewriting it as
\begin{align}
\vA^{\!\perp}(\vx)&=\vA(\vx)+\nabla\int\frac{\nabla^\prime\cdot
\vA(\vx^\prime)}{4\pi|\vx-\vx^\prime|}\,
d^3\vx^\prime\nonumber\\[0.5em]
& = \vA(\vx)+\nabla\int\nabla^\prime\cdot
\left(\frac{\vA(\vx^\prime)}{4\pi|\vx-\vx^\prime|}\right)\,
d^3\vx^\prime - \nabla\int
\vA(\vx^\prime)\cdot\left(\nabla^\prime\frac{1}{4\pi|\vx-\vx^\prime|}\right)\,
d^3\vx^\prime\nonumber\\[0.5em]
& = \vA(\vx) - \nabla\int
\vA(\vx^\prime)\cdot\left(\nabla^\prime\frac{1}{4\pi|\vx-\vx^\prime|}\right)\,
d^3\vx^\prime\nonumber\\[0.5em]
&= \vA(\vx)+  \nabla\int
\vA(\vx^\prime)\cdot\left(\nabla\frac{1}{4\pi|\vx-\vx^\prime|}\right)\,
d^3\vx^\prime\nonumber\\[0.5em]
&=\vA(\vx)+  \int
\vA(\vx^\prime)\cdot\left(\nabla\otimes\nabla\frac{1}{4\pi|\vx-\vx^\prime|}\right)\,
d^3\vx^\prime\nonumber\\
&= \int
d^3\vx^\prime\left(\delta^3(\vx-\vx^\prime)\cdot\boldsymbol{1}+
\nabla\otimes\nabla\frac{1}{4\pi|\vx-\vx^\prime|}\right)\cdot\vA(\vx^\prime),
\label{150809:1620}
\end{align}

\noindent where we made use of Gauss' theorem for the second term
in the second line, which resulted into a vanishing surface
integral due vanishing contributions of the vector field
$\vA(\vx)$ at infinity. Hence, the alternative relation
\eqref{150809:1620} for a transverse vector field allows to define
an operator $\boldsymbol{\mathcal{P}}$ of orthoprojection to
solenoidal vector fields:
$\vA\mapsto\boldsymbol{\mathcal{P}}[\vA]=\vA^{\!\perp}$. The
explicit expression for $\boldsymbol{\mathcal{P}}$ acting on any
arbitrary vector-field $\vA$ vanishing sufficiently fast at
infinity then has the form given by \eqref{150809:1620} (see e.g.
also Appendix B in~\cite{Frewer14.1}):
\begin{equation}
\boldsymbol{\mathcal{P}}[\vA]=\int d^3\vx^\prime
\vrho(\vx-\vx^\prime)\cdot \vA(\vx^\prime),
\end{equation}
with the kernel
\begin{equation}
\vrho(\vx-\vx^\prime):=\delta^3(\vx-\vx^\prime)\cdot\boldsymbol{1}+
\nabla\otimes\nabla\frac{1}{4\pi|\vx-\vx^\prime|}.
\end{equation}
By construction, the projection properties of
$\boldsymbol{\mathcal{P}}$ are
\begin{equation}
\left. \begin{aligned}
\boldsymbol{\mathcal{P}}[\vA]=\vA,\quad\text{if}\;\; \nabla\cdot
\vA=0,\hspace{1.4cm}\\
\boldsymbol{\mathcal{P}}[\vA]=\v0,\quad\text{if}\;\; \nabla\times
\vA=\v0, \;\;\text{e.g. if}\;\; \vA =\nabla\varphi .
\end{aligned}
~~~\right \}
\end{equation}

\section{The Hopf equation in two different representations}

If we assume that $\nabla\cdot\vy\neq 0$, for any arbitrary vector
field $\vy=\vy(\vx)$ which is decaying sufficiently fast at
infinity, the functional Hopf equation as given by
\eqref{150809:1945} can be represented in two alternative ways.
Either in a representation relative to the full vector field
$\vy$, or in a representation relative to its projected transverse
(solenoidal) part $\vy^\perp$.

\subsection{The Hopf equation in the full-field representation}

According to the defining relation \eqref{150809:2110}, the
functional Hopf equation \eqref{150809:1945} will take the form
\begin{align}
\frac{\partial \Phi}{\partial t} &=\int
\Big(y_\alpha(\vx)-\partial_\alpha\varphi(\vx)\Big)
\left(i\frac{\partial}{\partial x_\beta}\frac{\delta^2
\Phi}{\delta y_\alpha(\vx)\delta
y_\beta(\vx)}+\nu\Delta\frac{\delta\Phi}{\delta
y_\alpha(\vx)}\right)d^3\vx\nonumber\\[0.5em]
& = \int\! y_\alpha(\vx) \left(i\frac{\partial}{\partial
x_\beta}\frac{\delta^2 \Phi}{\delta y_\alpha(\vx)\delta
y_\beta(\vx)}+\nu\Delta\frac{\delta\Phi}{\delta
y_\alpha(\vx)}\right)d^3\vx\nonumber\\
&\qquad\qquad\quad +\int\! \varphi(\vx)
\left(i\frac{\partial^2}{\partial x_\alpha\partial
x_\beta}\frac{\delta^2 \Phi}{\delta y_\alpha(\vx)\delta
y_\beta(\vx)}+\nu\Delta\frac{\partial}{\partial
x_\alpha}\frac{\delta\Phi}{\delta
y_\alpha(\vx)}\right)d^3\vx\nonumber\\[0.5em]
& = \int\! y_\alpha(\vx) \left(i\frac{\partial}{\partial
x_\beta}\frac{\delta^2 \Phi}{\delta y_\alpha(\vx)\delta
y_\beta(\vx)}+\nu\Delta\frac{\delta\Phi}{\delta
y_\alpha(\vx)}\right)d^3\vx + i\!\int\! \varphi(\vx)
\frac{\partial^2}{\partial x_\alpha\partial x_\beta}\frac{\delta^2
\Phi}{\delta y_\alpha(\vx)\delta
y_\beta(\vx)}d^3\vx\nonumber\\[0.5em]
& = \int\! y_\alpha(\vx) \left(i\frac{\partial}{\partial
x_\beta}\frac{\delta^2 \Phi}{\delta y_\alpha(\vx)\delta
y_\beta(\vx)}+\nu\Delta\frac{\delta\Phi}{\delta
y_\alpha(\vx)}\right)d^3\vx\nonumber\\
&\qquad\qquad\quad +  i\!\int\!
y_\gamma(\vx^\prime)\left(\frac{\partial}{\partial
x^\prime_\gamma}\frac{1}{4\pi|\vx-\vx^\prime|}\right)\frac{\partial^2}{\partial
x_\alpha\partial x_\beta}\frac{\delta^2 \Phi}{\delta
y_\alpha(\vx)\delta y_\beta(\vx)}d^3\vx^\prime\,
d^3\vx\nonumber\\[0.5em]
& = \int\! y_\alpha(\vx) \left(i\frac{\partial}{\partial
x_\beta}\frac{\delta^2 \Phi}{\delta y_\alpha(\vx)\delta
y_\beta(\vx)}+\nu\Delta\frac{\delta\Phi}{\delta
y_\alpha(\vx)}\right)d^3\vx\nonumber\\
&\qquad\qquad\quad +  i\!\int\!
y_\alpha(\vx)\left(\frac{\partial}{\partial
x_\alpha}\frac{1}{4\pi|\vx-\vx^\prime|}\right)\frac{\partial^2}{\partial
x^\prime_\beta\partial x^\prime_\gamma}\frac{\delta^2 \Phi}{\delta
y_\beta(\vx^\prime)\delta y_\gamma(\vx^\prime)}d^3\vx^\prime\,
d^3\vx.
\label{150810:1006}
\end{align}

\noindent Note that in going from the second to the third
equality, the incompressibility condition in functional
space~\citep[p.~97]{Hopf52}\footnote[2]{Note that the
incompressibility condition only refers to the velocity field
$\vu$ of the flow, and {\it not} to the auxiliary field $\vy$. In
other words, although we assume $\nabla\cdot\vy\neq 0$, i.e.
$\vy\neq \vy^\perp$, we still have incompressibility in the
velocity field $\nabla\cdot \vu=0$, i.e. $\vu=\vu^\perp$; and in
functional space it is expressed as \eqref{150811:1348}.} has been
made used of

\begin{equation}
\frac{\partial}{\partial x_\alpha}\frac{\delta\Phi}{\delta
y_\alpha(\vx)}=0. \label{150811:1348}
\end{equation}

\subsection{The Hopf equation in the transverse representation}

In order to transform the Hopf equation \eqref{150809:1945} into a
representation where only the transverse fields $\vy^{\perp}$
appear, it is necessary to first transform the functional
derivatives according to relation \eqref{150809:2110} in applying
the corresponding functional chain rule
\begin{align}
\frac{\delta}{\delta y_\alpha (\vx)} & = \int d^3\vz\,
\frac{\delta y^\perp_\kappa(\vz)}{\delta
y_\alpha(\vx)}\frac{\delta}{\delta y^\perp_\kappa(\vz)},\\[0.5em]
\!\!\!\frac{\delta^2}{\delta y_\alpha (\vx)\delta y_\beta(\vx)} &
= \int d^3\vz\, \frac{\delta^2 y^\perp_\kappa(\vz)}{\delta
y_\alpha(\vx)\delta y_\beta(\vx)}\frac{\delta}{\delta
y^\perp_\kappa(\vz)} + \int d^3\vz\, d^3\vz^\prime\,\frac{\delta
y^\perp_\kappa(\vz)}{\delta y_\alpha(\vx)}\frac{\delta
y^\perp_\lambda(\vz^\prime)}{\delta
y_\beta(\vx)}\frac{\delta^2}{\delta y^\perp_\kappa(\vz)\delta
y^\perp_\lambda(\vz^\prime)}. \label{150810:1853}
\end{align}

\noindent Let's first consider the transformation of the single
functional derivative in the viscous term:
\begin{align}
\frac{\delta}{\delta y_\alpha (\vx)} & = \int
d^3\vz\left(\delta_{\alpha\kappa}\delta^3(\vz-\vx)+
\frac{\partial}{\partial z_\kappa}\int
\frac{\partial_\alpha^{\hspace{0.04cm}\prime}\,\delta^3(\vx^\prime-\vx)}{4\pi|\vz-\vx^\prime|}
d^3\vx^\prime \right)\frac{\delta}{\delta
y^\perp_\kappa(\vz)}\nonumber\\[0.5em]
& = \frac{\delta}{\delta y^\perp_\alpha (\vx)}-\int
d^3\vz\left(\frac{\partial}{\partial z_\kappa}\int
\delta^3(\vx^\prime-\vx)\frac{\partial}{\partial x^\prime_\alpha}
\frac{1}{4\pi|\vz-\vx^\prime|} d^3\vx^\prime
\right)\frac{\delta}{\delta y^\perp_\kappa(\vz)}\nonumber\\[0.5em]
& = \frac{\delta}{\delta y^\perp_\alpha (\vx)}-\int
d^3\vz\left(\frac{\partial}{\partial
z_\kappa}\frac{\partial}{\partial
x_\alpha}\frac{1}{4\pi|\vz-\vx|}\right)\frac{\delta}{\delta
y^\perp_\kappa(\vz)}\nonumber\\[0.5em]
& = \frac{\delta}{\delta y^\perp_\alpha (\vx)}+\int
d^3\vz\left(\frac{\partial}{\partial x_\alpha}
\frac{1}{4\pi|\vz-\vx|}\right)\frac{\partial}{\partial
z_\kappa}\frac{\delta}{\delta y^\perp_\kappa(\vz)}.
\label{150810:1834}
\end{align}

\noindent If this operator now acts on the characteristic
functional $\Phi$, which itself transforms invariantly
\cite[p.~93]{Hopf52}
\begin{equation}
\Phi[\vy(\vx),t]=\Phi[\vy^\perp(\vx),t],\label{150811:1500}
\end{equation}
then the second right-hand-side term in the last line of
\eqref{150810:1834} results to zero due to the incompressibility
condition \eqref{150811:1348}. Hence we obtain the following
invariant result
\begin{equation}
\frac{\delta\Phi[\vy(\vx),t]}{\delta y_\alpha (\vx^\prime)}=
\frac{\delta\Phi[\vy^\perp(\vx),t]}{\delta y^\perp_\alpha
(\vx^\prime)}.
\end{equation}
Similar for the second derivative in the inertial term, which also
gives the invariant result\footnote[2]{Note that since the
functional relation \eqref{150809:2110} for $\vy^\perp$ is linear
in $\vy$, the first term of the chain rule \eqref{150810:1853}
gives no contribution.}
\begin{align}
\frac{\delta^2\Phi}{\delta y_\alpha(\vx)\delta y_\beta(\vx)} & =
\int d^3\vz\, d^3\vz^\prime
\left(\delta_{\alpha\kappa}\delta^3(\vz-\vx)+
\frac{\partial}{\partial z_\kappa}\int
\frac{\partial_\alpha^{\hspace{0.04cm}\prime}\,
\delta^3(\vx^\prime-\vx)}{4\pi|\vz-\vx^\prime|}
d^3\vx^\prime \right)\nonumber\\
&\qquad\quad\;\;\, \cdot
\left(\delta_{\beta\lambda}\delta^3(\vz^\prime-\vx)+
\frac{\partial}{\partial z^\prime_\lambda}\int
\frac{\partial_\beta^{\hspace{0.04cm}\prime\prime}\,
\delta^3(\vx^{\prime\prime}-\vx)}{4\pi|\vz^\prime-\vx^{\prime\prime}|}
d^3\vx^{\prime\prime} \right)\frac{\delta^2\Phi}{\delta
y^\perp_\kappa(\vz)\delta
y^\perp_\lambda(\vz^\prime)}\nonumber\\[0.5em]
& = \int d^3\vz\, d^3\vz^\prime
\,\Bigg(\delta_{\alpha\kappa}\delta^3(\vz-\vx)-
\frac{\partial}{\partial z_\kappa}\frac{\partial}{\partial
x_\alpha}\frac{1}{4\pi|\vz-\vx|}\Bigg)\nonumber\\
&\qquad\quad\;\;\, \cdot
\left(\delta_{\beta\lambda}\delta^3(\vz^\prime-\vx)-
\frac{\partial}{\partial z^\prime_\lambda}\frac{\partial}{\partial
x_\beta} \frac{1}{4\pi|\vz^\prime-\vx|}
\right)\frac{\delta^2\Phi}{\delta y^\perp_\kappa(\vz)\delta
y^\perp_\lambda(\vz^\prime)}\nonumber
\end{align}
\begin{align}
& = \frac{\delta^2\Phi}{\delta
y^\perp_\alpha(\vx)\delta y^\perp_\beta(\vx)}\nonumber\\
& \quad\: + \int d^3\vz\, d^3\vz^\prime\,
\delta_{\alpha\kappa}\delta^3(\vz-\vx)\frac{\partial}{\partial
x_\beta} \frac{1}{4\pi|\vz^\prime-\vx|}\cdot\frac{\delta}{\delta
y^\perp_\kappa(\vz)}\left(\frac{\partial}{\partial
z^\prime_\lambda}\frac{\delta\Phi}{\delta
y^\perp_\lambda(\vz^\prime)}\right)\nonumber\\
& \quad\: + \int d^3\vz\,
d^3\vz^\prime\,\delta_{\beta\lambda}\delta^3(\vz^\prime-\vx)\frac{\partial}{\partial
x_\alpha} \frac{1}{4\pi|\vz-\vx|}\cdot\frac{\delta}{\delta
y^\perp_\lambda(\vz^\prime)}\,\Bigg(\frac{\partial}{\partial
z_\kappa}\frac{\delta\Phi}{\delta
y^\perp_\kappa(\vz)}\Bigg)\nonumber\\
& \quad\: + \int d^3\vz\,
d^3\vz^\prime\,\Bigg(\frac{\partial}{\partial
x_\alpha}\frac{1}{4\pi|\vz-\vx|}\Bigg)\left(\frac{\partial}{\partial
x_\beta}
\frac{1}{4\pi|\vz^\prime-\vx|}\right)\cdot\frac{\partial}{\partial
z_\kappa}\frac{\delta}{\delta
y^\perp_\kappa(\vz)}\left(\frac{\partial}{\partial
z^\prime_\lambda}\frac{\delta\Phi}{\delta
y^\perp_\lambda(\vz^\prime)}\right)\nonumber\\[0.5em]
& = \frac{\delta^2\Phi}{\delta y^\perp_\alpha(\vx)\delta
y^\perp_\beta(\vx)},
\end{align}

\noindent where again we made use of the incompressibility
condition \eqref{150811:1348}. Now, since the characteristic
functional $\Phi$ transforms invariantly \eqref{150811:1500}, the
transformed Hopf equation \eqref{150809:1945} in the transverse
representation finally takes the form
\begin{align}
\frac{\partial \Phi}{\partial t} & = \int y^{\perp}_\alpha(\vx)
\left(i\frac{\partial}{\partial x_\beta}\frac{\delta^2
\Phi}{\delta y^\perp_\alpha(\vx)\delta
y^\perp_\beta(\vx)}+\nu\Delta\frac{\delta\Phi}{\delta
y^\perp_\alpha(\vx)}\right)d^3\vx. \label{150811:1503}
\end{align}

\section{The MPC equations in two different representations}

Similar as in the case for the Hopf equation (see Appendix B), the
MPC equations for the incompressible Navier-Stokes velocity field
$\vu=\vu(\vx,t)$ can be represented in two alternative ways too.
Again, either in a representation relative to the full MPC
fields\footnote[2]{The $H$-notation for the (instantaneous) MPC
functions in \eqref{150811:2236} is taken from \cite{Oberlack10};
see also \cite{Frewer14.1} for a critical discussion on using this
notation, in particular for invariance analysis.}
\begin{equation}
H_{\alpha_{\{n\}}}:=H_{\alpha_1\dotsc\alpha_n}:=\big\L
u_{\alpha_1}(\vx_1)\cdots u_{\alpha_n}(\vx_n)\big\R,
\label{150811:2236}
\end{equation}
or in a representation relative to the transversely projected MPC
fields
\begin{equation}
H^\perp_{\alpha_{\{n\}}}:=H^\perp_{\alpha_1\dotsc\alpha_n}:=\big\L
u_{\alpha_1}(\vx_1)\cdots u_{\alpha_n}(\vx_n)\big\R^\perp.
\end{equation}
Note that due to \eqref{150811:2104}, along with its evaluation
\eqref{150811:2353}, we have the obvious invariant relation
$H^\perp_{\alpha_{\{n\}}}=H_{\alpha_{\{n\}}}$. This invariance can
be readily verified independently from relation
\eqref{150811:2104}, by recognizing that i) the (fluctuating)
velocity field is solenoidal
$\partial_{\alpha_k}u_{\alpha_k}(\vx_k)=0$, for any index~$k$,
i.e. $u_{\alpha_k}(\vx_k)=u^\perp_{\alpha_k}(\vx_k)$, and ii) that
any usual (non-functional) differential operator acting locally on
any point $\vx_k$ is commuting with the averaging (ensemble)
operator $\L\cdot\R$. Hence we have the relation
\begin{gather}
\partial_{\alpha_k}\big\L
u_{\alpha_1}(\vx_1)\cdots u_{\alpha_k}(\vx_k)\cdots
u_{\alpha_n}(\vx_n)\big\R = \big\L u_{\alpha_1}(\vx_1)\cdots
\partial_{\alpha_k}u_{\alpha_k}(\vx_k)\cdots
u_{\alpha_n}(\vx_n)\big\R=0\\[0.5em]
\Longleftrightarrow\qquad\;\;\;\;\phantom{x}\nonumber\\[0.25em]
\big\L u_{\alpha_1}(\vx_1)\cdots u_{\alpha_n}(\vx_n)\big\R=\big\L
u_{\alpha_1}(\vx_1)\cdots
u_{\alpha_n}(\vx_n)\big\R^\perp.\qquad\:\phantom{x}
\label{150812:0058}
\end{gather}

\noindent But careful: When taking the smooth and regular limit of
a zero-correlation length between any two
points~$|\vx_k-\vx_l|\rightarrow 0$, the invariance
\eqref{150812:0058} no longer holds, because obviously
\begin{multline}
\partial_{\alpha_k}\lim_{\vx_l\rightarrow\vx_k}\big\L
u_{\alpha_1}(\vx_1)\cdots u_{\alpha_k}(\vx_k)\cdots
u_{\alpha_l}(\vx_l)\cdots u_{\alpha_n}(\vx_n)\big\R\\
=\,\,\big\L u_{\alpha_1}(\vx_1)\cdots u_{\alpha_k}(\vx_k)\cdots
\partial_{\alpha_k} u_{\alpha_l}(\vx_k)\cdots
u_{\alpha_n}(\vx_n)\big\R\,\, \neq\,\, 0. \label{150812:2116}
\end{multline}

\subsection{The MPC equations in the transverse representation}

The underlying MPC-generating equation is given by
\eqref{150812:0807}
\begin{equation}
\frac{\partial \Phi^n}{\partial t} = \int y^{\perp}_\alpha(\vx)
\left(i\frac{\partial}{\partial x_\beta}\frac{\delta^2
\Phi^{n+1}}{\delta y^\perp_\alpha(\vx)\delta
y^\perp_\beta(\vx)}+\nu\Delta\frac{\delta\Phi^n}{\delta
y^\perp_\alpha(\vx)}\right)d^3\vx, \quad \forall n\geq 1,
\label{150812:0808}
\end{equation}
where the hierarchy of functionals $\Phi^n$ is given by
\eqref{150812:0812}, along with \eqref{150811:2104} and
\eqref{150811:2353}. Three terms are contributing to equation
\eqref{150812:0808}: the temporal term on the left-hand side and
the inertial and viscous terms on the right-hand side, which for
each order $n$ evaluates to (shown here only up third order)
\begin{align}
n=1:\quad\; & \frac{\partial\Phi^1}{\partial t}=\int d^3\vx_1\,
y^\perp_{\alpha_1}(\vx_1)\left(i\,\frac{\partial}{\partial t}
\big\L
u_{\alpha_1}(\vx_1)\big\R^\perp\right),\\[0.5em]
& \frac{\delta^2 \Phi^{2}}{\delta y^\perp_\alpha(\vx)\delta
y^\perp_\beta(\vx)}=-\frac{2\cdot 1}{2!}\,
\big\L u_{\alpha}(\vx) u_{\beta}(\vx)\big\R^\perp,\\[0.5em]
& \frac{\delta\Phi^1}{\delta y^\perp_\alpha(\vx)}=i\,\big\L
u_{\alpha}(\vx)\big\R^\perp,
\\[1.5em]
n=2:\quad\; & \frac{\partial\Phi^2}{\partial t}=\int d^3\vx_1\,
d^3\vx_2\, y^\perp_{\alpha_1}(\vx_1)\,y^\perp_{\alpha_2}(\vx_2)
\left(-\frac{1}{2!}\,\frac{\partial}{\partial t} \big\L
u_{\alpha_1}(\vx_1)
u_{\alpha_2}(\vx_2)\big\R^\perp\right),\label{150812:1521}\\[0.5em]
& \frac{\delta^2 \Phi^{3}}{\delta y^\perp_\alpha(\vx)\delta
y^\perp_\beta(\vx)}=\int d^3\vx_3\,
y_{\alpha_3}^\perp(\vx_3)\left(-i\,\frac{3\cdot 2}{3!}\,\big\L
u_{\alpha}(\vx) u_{\beta}(\vx)
u_{\alpha_3}(\vx_3)\big\R^\perp\right),\label{150812:1514}\\[0.5em]
& \frac{\delta\Phi^2}{\delta y^\perp_\alpha(\vx)} =\int d^3\vx_2\,
y_{\alpha_2}^\perp(\vx_2)\left(-\frac{2}{2!}\,\big\L
u_{\alpha}(\vx) u_{\alpha_2}(\vx_2)\big\R^\perp\right),
\\[1.5em]
n=3:\quad\; & \frac{\partial\Phi^3}{\partial t}=\int d^3\vx_1\,
d^3\vx_2\,d^2\vx_3\,
y^\perp_{\alpha_1}(\vx_1)\,y^\perp_{\alpha_2}(\vx_2)\,
y^\perp_{\alpha_3}(\vx_3)\nonumber\\
&\text{\hspace{4cm}}\cdot\left(-i\,\frac{1}{3!}\,\frac{\partial}{\partial
t} \big\L u_{\alpha_1}(\vx_1)
u_{\alpha_2}(\vx_2)u_{\alpha_3}(\vx_3)\big\R^\perp\right),\\[0.5em]
& \frac{\delta^2 \Phi^{4}}{\delta y^\perp_\alpha(\vx)\delta
y^\perp_\beta(\vx)}=\int d^3\vx_3\,d^3\vx_4\,
y_{\alpha_3}^\perp(\vx_3)y_{\alpha_4}^\perp(\vx_4)\nonumber\\
&\text{\hspace{4.8cm}}\cdot\left(\frac{4\cdot 3}{4!}\,\big\L
u_{\alpha}(\vx) u_{\beta}(\vx)
u_{\alpha_3}(\vx_3)u_{\alpha_4}(\vx_4)\big\R^\perp\right),\\[0.5em]
& \frac{\delta\Phi^3}{\delta y^\perp_\alpha(\vx)} =\int
d^3\vx_2\,d^3\vx_3\,
y_{\alpha_2}^\perp(\vx_2)y_{\alpha_3}^\perp(\vx_3)\left(
-i\,\frac{3}{3!}\,\big\L u_{\alpha}(\vx)
u_{\alpha_2}(\vx_2)u_{\alpha_3}(\vx_3)\big\R^\perp\right),
\end{align}

\noindent and so on for all higher orders. When inserting these
results for each order into equation \eqref{150812:0808}, two
things should be noted or to be carried out before dropping the
integrals in order to obtain the MPC equations in the transverse
representation: i) The integrals for the inertial and viscous
terms must be fully symmetrized in their integration variables in
accord with the integrals from the temporal term, e.g. consider
the second order result for the inertial term~\eqref{150812:1514},
which, through equation \eqref{150812:0808} and in accord with
result \eqref{150812:1521}, has to be symmetrized to
\begin{multline}
\int d^3\vx\,d^3\vx_3\,
y^\perp_\alpha(\vx)\,y^\perp_{\alpha_3}(\vx_3)\,\frac{\partial}{\partial
x_\beta}\,\big\L u_{\alpha}(\vx) u_{\beta}(\vx)
u_{\alpha_3}(\vx_3)\big\R^\perp\\
=\int d^3\vx_1\, d^3\vx_2\,
y^\perp_{\alpha_1}(\vx_1)\,y^\perp_{\alpha_2}(\vx_2)\Bigg(\;\;\,\frac{1}{2}\,
\frac{\partial}{\partial x_{1,\sigma_1}}\,\big\L
u_{\alpha_1}(\vx_1)
u_{\alpha_2}(\vx_2) u_{\sigma_1}(\vx_1)\big\R^\perp\\
\phantom{x}\text{\hspace{5.5cm}}+\,\frac{1}{2}\,
\frac{\partial}{\partial x_{2,\sigma_2}}\,\big\L
u_{\alpha_1}(\vx_1) u_{\alpha_2}(\vx_2)
u_{\sigma_2}(\vx_2)\big\R^\perp\;\,\Bigg)\\
=\int d^3\vx_1\, d^3\vx_2\,
y^\perp_{\alpha_1}(\vx_1)\,y^\perp_{\alpha_2}(\vx_2)
\left(\,\frac{1}{2}\,\frac{\partial}{\partial x_{1,\sigma_1}}\,
\widehat{H}^\perp_{\alpha_1\alpha_2\sigma_1}
+\frac{1}{2}\,\frac{\partial}{\partial x_{2,\sigma_2}}\,
\widehat{H}^\perp_{\alpha_1\alpha_2\sigma_2}\right),
\text{\hspace{0.0cm}} \label{150812:1920}
\end{multline}
and ii) the invariance condition $\vH^\perp_n=\vH_n$
\eqref{150812:0058} is only valid if the MPC function of $n$-th
order $\vH^\perp_n=(H^\perp_{\alpha_1\dotsc\alpha_n})$ is also
evaluated at $n$ distinct points $\vx_n$. As soon as it's
evaluated in less than $n$ distinct points, the invariance no
longer holds \eqref{150812:2116}. These lower-dimensional MPC
functions will be denoted by $\widehat{\vH}^\perp_n$, as it was
already done in \eqref{150812:1920}, in order to avoid any
uncertainties in the notation. For them, as already said, the
invariance condition does not hold, i.e.
$\widehat{\vH}^\perp_n\neq\widehat{\vH}_n$. Following both these
instructions i) and ii), we finally obtain the hierarchy of MPC
equations in the transverse representation (shown here only up
third order)
\begin{align}
n=1:\quad\; & \frac{\partial H_{\alpha_1}}{\partial t}
+\frac{\partial\widehat{H}^\perp_{\alpha_1\sigma_1}}{\partial
x_{1,\sigma_1}}-\nu\frac{\partial^2 H_{\alpha_1}}{\partial
x_{1,\sigma_1}\partial x_{1,\sigma_1}}=0,
\label{150813:0800}\\[1em]
n=2:\quad\; & \frac{\partial H_{\alpha_1\alpha_2}}{\partial t} +
\frac{\partial\widehat{H}^\perp_{\alpha_1\alpha_2\sigma_1}}{\partial
x_{1,\sigma_1}}+\frac{\partial\widehat{H}^\perp_{\alpha_1\alpha_2\sigma_2}}{\partial
x_{2,\sigma_2}}-\nu\frac{\partial^2 H_{\alpha_1\alpha_2}}{\partial
x_{1,\sigma_1}\partial x_{1,\sigma_1}}-\nu\frac{\partial^2
H_{\alpha_1\alpha_2}}{\partial x_{2,\sigma_2}\partial
x_{2,\sigma_2}}=0,
\\[1em]
n=3:\quad\; & \frac{\partial
H_{\alpha_1\alpha_2\alpha_3}}{\partial t}+
\frac{\partial\widehat{H}^\perp_{\alpha_1\alpha_2\alpha_3\sigma_1}}{\partial
x_{1,\sigma_1}}+
\frac{\partial\widehat{H}^\perp_{\alpha_1\alpha_2\alpha_3\sigma_2}}{\partial
x_{2,\sigma_2}}+
\frac{\partial\widehat{H}^\perp_{\alpha_1\alpha_2\alpha_3\sigma_3}}{\partial
x_{3,\sigma_3}}\nonumber\\
&\text{\hspace{2.8cm}} -\nu\frac{\partial^2
H_{\alpha_1\alpha_2\alpha_3}}{\partial x_{1,\sigma_1}\partial
x_{1,\sigma_1}}-\nu\frac{\partial^2
H_{\alpha_1\alpha_2\alpha_3}}{\partial x_{2,\sigma_2}\partial
x_{2,\sigma_2}}-\nu\frac{\partial^2
H_{\alpha_1\alpha_2\alpha_3}}{\partial x_{3,\sigma_3}\partial
x_{3,\sigma_3}}=0,\label{150813:0801}
\end{align}

\noindent where the coupling within the hierarchy is given through
the definition
\begin{equation}
\widehat{H}^\perp_{\alpha_1\dotsc\alpha_n\sigma_k}=\left(
\lim_{\vx_{n+1}\to\vx_k;\,\alpha_{n+1}\to\sigma_k}
H_{\alpha_1\dotsc\alpha_n\alpha_{n+1}}\right)^{\!\perp}\!
,\;\;\:\text{$\forall k$ between $1\leq k\leq n$}.
\end{equation}
Note that the infinite hierarchy of MPC equations is accompanied
by the following obvious incompressibility constraints
\begin{gather}
\frac{\partial H_{\alpha_1\cdots\alpha_n}}{\partial
x_{k,\alpha_k}}=0,
\;\;\:\text{$\forall k$ between $1\leq k\leq n$},\\
\frac{\partial\widehat{H}^\perp_{\alpha_1\cdots\alpha_n\alpha_l}}{\partial
x_{k,\alpha_k}}=0, \;\;\:\text{$\forall k,l$ between $1\leq
(k,l)\leq n$, for $k\neq l$.} \label{150813:1259}
\end{gather}

\subsection{The MPC equations in the full-field representation}

The full-field representation of the MPC equations follows
directly from \eqref{150813:0800}-\eqref{150813:0801} by
decom\-posing each of the inertial transverse terms according to
relation \eqref{150809:2110}. But, as written, relation
\eqref{150809:2110} only applies for vector fields. Therefore it
is necessary to first extend this definition to also include
tensor fields of arbitrary rank such that it applies to
\eqref{150813:0800}-\eqref{150813:0801}. Following the principle
of \eqref{150809:2110} it can straightforwardly be generalized to
construct transverse (solenoidal) tensor fields
\begin{equation}
A^{\perp}_{\alpha_1\cdots\alpha_n}(\vx)=A_{\alpha_1\cdots\alpha_n}(\vx)
+\frac{1}{n}\sum_{k=1}^n
\partial_{\alpha_k}\int
\frac{\partial_{\alpha^\prime_k}^{\hspace{0.04cm}\prime}
A_{\alpha_1\cdots \alpha^\prime_k\cdots \alpha_n}(\vx^\prime)}
{4\pi|\vx-\vx^\prime|}\, d^3\vx^\prime, \label{150813:0904}
\end{equation}
where $\vA_n=(A_{\alpha_1\cdots\alpha_n})$ is an arbitrary tensor
field of rank $n$ decaying sufficiently fast at infinity. By
construction, \eqref{150813:0904} is in accord with
\eqref{150809:2110} since its full-ranked divergence
vanishes\footnote[2]{Note that only a full ranked divergence on
$A^{\perp}_{\alpha_1\cdots\alpha_n}$ \eqref{150813:0904} vanishes,
i.e. when all indices are contracted. A partial divergence where
not all indices are contracted does not result to zero. Hence, the
partial divergences of the transverse MPC fields appearing in
\eqref{150813:0800}-\eqref{150813:0801} are non-zero.}
\begin{equation}
\partial^{\,
n}_{\alpha_1\cdots\alpha_n}A^{\perp}_{\alpha_1\cdots\alpha_n}(\vx)=0.
\end{equation}
It is obvious that relation \eqref{150813:0904} also further
extends to {\it multi}-point tensor fields, hence we obtain the
following decompositions of the transverse MPC fields appearing in
\eqref{150813:0800}-\eqref{150813:0801} (shown here only up to
second order)
\begin{align}
n=1:\quad\; &
\frac{\partial\widehat{H}^\perp_{\alpha_1\sigma_1}(\vx_1)}{\partial
x_{1,\sigma_1}}=\frac{\partial\widehat{H}_{\alpha_1\sigma_1}(\vx_1)}{\partial
x_{1,\sigma_1}}+\frac{\partial}{\partial x_{1,\alpha_1}} \int
\frac{d^3\vx_1^\prime}{4\pi|\vx_1-\vx_1^\prime|}\frac{\partial^2
\widehat{H}_{\alpha_1^\prime\sigma_1^\prime}(\vx_1^\prime)}{\partial
x^\prime_{1,\alpha_1^\prime}\partial
x^\prime_{1,\sigma_1^\prime}}\nonumber\\[0.5em]
&\text{\hspace{2cm}} =
\frac{\partial\widehat{H}_{\alpha_1\sigma_1}(\vx_1)}{\partial
x_{1,\sigma_1}}+\frac{\partial}{\partial x_{1,\alpha_1}} \int
\frac{d^3\vx_1^\prime}{4\pi|\vx_1-\vx_1^\prime|}\frac{\partial^2
\big\L
u_{\alpha_1^\prime}(x_1^\prime)u_{\sigma_1^\prime}(x_1^\prime)\big\R}{\partial
x^\prime_{1,\alpha_1^\prime}\partial x^\prime_{1,\sigma_1^\prime}}\nonumber\\[0.5em]
& \text{\hspace{2cm}} =
\frac{\partial\widehat{H}_{\alpha_1\sigma_1}(\vx_1)}{\partial
x_{1,\sigma_1}} +\frac{\partial \big\L p(\vx_1)\big\R}{\partial
x_{1,\alpha_1}}\,\, =:\,\,
\frac{\partial\widehat{H}_{\alpha_1\sigma_1}(\vx_1)}{\partial
x_{1,\sigma_1}} +\frac{\partial I_{[1]}(\vx_1)}{\partial
x_{1,\alpha_1}},
\\[1.5em]
n=2:\quad\; &
\frac{\partial\widehat{H}^\perp_{\alpha_1\alpha_2\sigma_1}(\vx_1,\vx_2)}{\partial
x_{1,\sigma_1}}+\frac{\partial\widehat{H}^\perp_{\alpha_1\alpha_2\sigma_2}(\vx_1,\vx_2)}
{\partial x_{2,\sigma_2}}\nonumber\\[0.5em]
&\text{\hspace{0.5cm}} = \;\;\;\;\;
\frac{\partial\widehat{H}_{\alpha_1\alpha_2\sigma_1}(\vx_1,\vx_2)}{\partial
x_{1,\sigma_1}} + \frac{1}{2}\,\frac{\partial}{\partial
x_{1,\alpha_1}} \int
\frac{d^3\vx_1^\prime}{4\pi|\vx_1-\vx_1^\prime|}\frac{\partial^2
\widehat{H}_{\alpha_1^\prime\alpha_2\sigma_1^\prime}(\vx_1^\prime,\vx_2)}{\partial
x^\prime_{1,\alpha_1^\prime}\partial
x^\prime_{1,\sigma_1^\prime}}\nonumber\\[0.5em]
&\text{\hspace{4.5cm}} + \frac{1}{2}\,\frac{\partial}{\partial
x_{2,\alpha_2}} \int
\frac{d^3\vx_2^\prime}{4\pi|\vx_2-\vx_2^\prime|}\frac{\partial^2
\widehat{H}_{\alpha_1\alpha_2^\prime\sigma_2^\prime}(\vx_1,\vx_2^\prime)}{\partial
x^\prime_{2,\alpha_2^\prime}\partial
x^\prime_{2,\sigma_2^\prime}}\nonumber\\[0.5em]
&\text{\hspace{1.0cm}}
+\;\,\frac{\partial\widehat{H}_{\alpha_1\alpha_2\sigma_2}(\vx_1,\vx_2)}
{\partial x_{2,\sigma_2}} + \frac{1}{2}\,\frac{\partial}{\partial
x_{2,\alpha_2}} \int
\frac{d^3\vx_2^\prime}{4\pi|\vx_2-\vx_2^\prime|}\frac{\partial^2
\widehat{H}_{\alpha_1\alpha_2^\prime\sigma_2^\prime}(\vx_1,\vx_2^\prime)}{\partial
x^\prime_{2,\alpha_2^\prime}\partial
x^\prime_{2,\sigma_2^\prime}}\nonumber\\[0.5em]
&\text{\hspace{4.5cm}} + \frac{1}{2}\,\frac{\partial}{\partial
x_{1,\alpha_1}} \int
\frac{d^3\vx_1^\prime}{4\pi|\vx_1-\vx_1^\prime|}\frac{\partial^2
\widehat{H}_{\alpha_1^\prime\alpha_2\sigma_1^\prime}(\vx_1^\prime,\vx_2)}{\partial
x^\prime_{1,\alpha_1^\prime}\partial
x^\prime_{1,\sigma_1^\prime}}\nonumber\\[0.5em]
& =\,
\frac{\partial\widehat{H}_{\alpha_1\alpha_2\sigma_1}(\vx_1,\vx_2)}{\partial
x_{1,\sigma_1}}+\frac{\partial \big\L
p(\vx_1)u_{\alpha_2}(\vx_2)\big\R}{\partial x_{1,\alpha_1}}
+\frac{\partial\widehat{H}_{\alpha_1\alpha_2\sigma_2}(\vx_1,\vx_2)}
{\partial x_{2,\sigma_2}}+\frac{\partial \big\L
u_{\alpha_1}(\vx_1)p(\vx_2)\big\R}{\partial
x_{2,\alpha_2}}\nonumber\\[0.5em]
&  =:\,
\frac{\partial\widehat{H}_{\alpha_1\alpha_2\sigma_1}(\vx_1,\vx_2)}{\partial
x_{1,\sigma_1}}+\frac{\partial I_{\alpha_2
[1]}(\vx_1,\vx_2)}{\partial
x_{1,\alpha_1}}+\frac{\partial\widehat{H}_{\alpha_1\alpha_2\sigma_2}(\vx_1,\vx_2)}
{\partial x_{2,\sigma_2}}+\frac{\partial I_{\alpha_1
[2]}(\vx_1,\vx_2)}{\partial x_{2,\alpha_2}},
\end{align}

\noindent wich then according to
\eqref{150813:0800}-\eqref{150813:0801} finally results into the
infinite hierarchy of MPC equations in the full-field
representation (shown here only up to third order)
\begin{align}
n=1:\quad\; & \frac{\partial H_{\alpha_1}}{\partial t}
+\frac{\partial\widehat{H}_{\alpha_1\sigma_1}}{\partial
x_{1,\sigma_1}}+\frac{\partial I_{[1]}}{\partial
x_{1,\alpha_1}}-\nu\frac{\partial^2 H_{\alpha_1}}{\partial
x_{1,\sigma_1}\partial x_{1,\sigma_1}}=0,
\label{150813:1241}\\[1em]
n=2:\quad\; & \frac{\partial H_{\alpha_1\alpha_2}}{\partial t} +
\frac{\partial\widehat{H}_{\alpha_1\alpha_2\sigma_1}}{\partial
x_{1,\sigma_1}}+\frac{\partial\widehat{H}_{\alpha_1\alpha_2\sigma_2}}{\partial
x_{2,\sigma_2}}+\frac{\partial I_{\alpha_2 [1]}}{\partial
x_{1,\alpha_1}}+\frac{\partial I_{\alpha_1 [2]}}{\partial
x_{2,\alpha_2}}\nonumber\\
&\text{\hspace{5.5cm}}-\nu\frac{\partial^2
H_{\alpha_1\alpha_2}}{\partial x_{1,\sigma_1}\partial
x_{1,\sigma_1}}-\nu\frac{\partial^2 H_{\alpha_1\alpha_2}}{\partial
x_{2,\sigma_2}\partial x_{2,\sigma_2}}=0,
\\[1em]
n=3:\quad\; & \frac{\partial
H_{\alpha_1\alpha_2\alpha_3}}{\partial t}+
\frac{\partial\widehat{H}_{\alpha_1\alpha_2\alpha_3\sigma_1}}{\partial
x_{1,\sigma_1}}+
\frac{\partial\widehat{H}_{\alpha_1\alpha_2\alpha_3\sigma_2}}{\partial
x_{2,\sigma_2}}+
\frac{\partial\widehat{H}_{\alpha_1\alpha_2\alpha_3\sigma_3}}{\partial
x_{3,\sigma_3}}\nonumber\\
&\text{\hspace{2.8cm}} +\frac{\partial I_{\alpha_2\alpha_3
[1]}}{\partial x_{1,\alpha_1}}+\frac{\partial I_{\alpha_1\alpha_3
[2]}}{\partial x_{2,\alpha_2}}+\frac{\partial I_{\alpha_1\alpha_2
[3]}}{\partial
x_{3,\alpha_3}}\nonumber\\
&\text{\hspace{2.8cm}} -\nu\frac{\partial^2
H_{\alpha_1\alpha_2\alpha_3}}{\partial x_{1,\sigma_1}\partial
x_{1,\sigma_1}}-\nu\frac{\partial^2
H_{\alpha_1\alpha_2\alpha_3}}{\partial x_{2,\sigma_2}\partial
x_{2,\sigma_2}}-\nu\frac{\partial^2
H_{\alpha_1\alpha_2\alpha_3}}{\partial x_{3,\sigma_3}\partial
x_{3,\sigma_3}}=0.\label{150813:1242}
\end{align}

\noindent As before in the transverse field representation
\eqref{150813:0800}-\eqref{150813:1259}, the above equations are
coupled through the definition
\begin{equation}
\widehat{H}_{\alpha_1\dotsc\alpha_n\sigma_k}=
\lim_{\vx_{n+1}\to\vx_k;\,\alpha_{n+1}\to\sigma_k}
H_{\alpha_1\dotsc\alpha_n\alpha_{n+1}},\;\;\:\text{$\forall k$
between $1\leq k\leq n$}, \label{150813:1514}
\end{equation}
and are accompanied by the incompressibility constraints
\begin{gather}
\frac{\partial H_{\alpha_1\cdots\alpha_n}}{\partial
x_{k,\alpha_k}}=0,
\;\;\:\text{$\forall k$ between $1\leq k\leq n$},\\
\frac{\partial
I_{\alpha_1\cdots\alpha_{l-1}\alpha_{l+1}\cdots\alpha_n
[l]}}{\partial x_{k,\alpha_k}}=0, \;\;\:\text{$\forall k,l$
between $1\leq (k,l)\leq n$, for $k\neq l$.} \label{150813:1305}
\end{gather}
Note that the above system agrees with the MPC hierarchy of
equations derived in \cite{Oberlack10}, from which also the above
notation has been borrowed (see also
\cite{Frewer14.1,Frewer14.3}). These are i) the instantaneous
(equal-time) multi-point velocity correlation functions of order
$n\geq 1$
\begin{equation}
H_{\alpha_{\{n\}}}=H_{\alpha_1\cdots\alpha_n}(\vx_1,\dotsc,\vx_n,t)=\big\L
u_{\alpha_1}(\vx_1,t)\cdots u_{\alpha_n}(\vx_n,t)\big\R,
\end{equation}
where
$\widehat{H}_{\alpha_1\cdots\alpha_n\sigma_k}(\vx_1,\dotsc,\vx_n,\vx_k,t)$
is a lower dimensional moment which emerges from the next higher
one
$H_{\alpha_1\cdots\alpha_n\alpha_{n+1}}(\vx_1,\dotsc,\vx_n,\vx_{n+1},t)$
in the limit of zero correlation length
$|\vx_{n+1}-\vx_k|\rightarrow 0$ along with a corresponding change
in the tensor index $\alpha_{n+1}\rightarrow \sigma_k$, for all
$k$ within the range $1\leq k\leq n$ \eqref{150813:1514}, and then
ii) the instantaneous (equal-time) multi-point pressure velocity
correlation functions of order $(n-1)$
\begin{equation}
I_{\alpha_{\{n-1\}}[l]}=\big\L u_{\alpha_1}(\vx_1,t)\cdots
u_{\alpha_{l-1}}(\vx_{l-1},t)\cdot p(\vx_l,t)\cdot
u_{\alpha_{l+1}}(\vx_{l+1},t)\cdots u_{\alpha_n}(\vx_n,t)\big\R,
\end{equation}
where $p$ is the instantaneous pressure field, which in the
unbounded domain evolves according to Poisson's equation as (see
e.g. \cite{McComb90})
\begin{equation}
p(\vx,t)=\int
d^3\vx^\prime\,\frac{\partial^{\hspace{0.03cm}\prime\hspace{0.02cm}
2}_{\alpha\beta} u_\alpha(\vx^\prime,t)u_\beta(\vx^\prime,t)}
{4\pi|\vx-\vx^\prime|}.
\end{equation}

\section{Derivation of the ODE moment equations from the
underlying PDE}

In a similar fashion as the infinite (lower level)
Friedmann-Keller hierarchy of moments \eqref{150817:1647} emerge
from the underlying (higher level) Hopf equation
\eqref{150811:1223}, the analogous lower level ODE moment
equations \eqref{150817:2057} are induced by the higher level PDE
\eqref{150817:2056} in the following way: Multiplying the
unbounded PDE for $u=u(x,t)$
\begin{equation}
\partial_t u = \partial_x^2 u-\lambda\cdot x^2\, u,\label{150826:1107}
\end{equation}
by $x^n$, for all $n\geq 0$, and integrating over $\mathbb{R}$ by
assuming that partial integration is justified, i.e. by assuming
the natural boundary conditions
\begin{equation}
\lim_{x\to \pm \infty} u(x,t)=0,\quad\;
\lim_{x\to\pm\infty}\partial_x u(x,t)=0,\quad\;\forall t\geq 0,
\end{equation}
then one obtains for each order $n$ the following infinite system
of coupled ODEs
\begin{equation}
\frac{d u_n}{dt}=n\cdot (n-1)\cdot u_{n-2}-\lambda\cdot
u_{n+2},\;\;\; \forall n\geq 0,\label{150826:1113}
\end{equation}
where the moments $u_n=u_n(t)$ are defined by
\begin{equation}
u_n(t)=\int_{-\infty}^\infty x^n\cdot u(x,t)\, dx,\;\; \forall
n\geq 0.\label{150826:1117}
\end{equation}
Formulating \eqref{150826:1107} as a Cauchy problem, by including
the initial condition
\begin{equation}
u(x,0)=\phi(x),\;\;\text{with}\;\; \int_{-\infty}^\infty\phi(x)\,
dx=1,
\end{equation}
where $\phi$ is some arbitrary but integrable function, it will
correspondingly restrict the infinite system of equations
\eqref{150826:1113} to a first order ODE initial value problem, in
that it has to satisfy the restrictions
\begin{equation}
u_n(0)=\int_{-\infty}^\infty x^n\cdot\phi(x)\, dx,\;\; \forall
n\geq 0,\;\;\text{with}\;\; u_0(0)=1.
\end{equation}

\bibliographystyle{jfm}
\bibliography{BibDaten}

\end{document}